\RequirePackage{fix-cm}
\documentclass[natbib,smallextended]{svjour3}       
\smartqed  
\usepackage{graphicx}
\usepackage{hyperref}
\usepackage{amssymb,amsmath}

 \journalname{my journal}
\def\ompt{\omega_{\rm p}t}
\def\comp{\,c/\omega_{\rm p}}

\newcommand{\fig}[1]{Fig.~\ref{fig:#1}}
\newcommand{\bmath}[1]{\mbox{\boldmath{$#1$}}}
\newcommand{\pasp}{{Publ. Astron. Soc. Pac.}}

\newcommand{\nat}{{Nature}}

\newcommand{\aap}{{Astron. Astrophys.}}
\newcommand{\mnras}{{Mon. Not. R. Astron. Soc.}}

\newcommand{\apj}{{Astrophys. J.}}
\newcommand{\aj}{{Astron. J.}}
\newcommand{\apjl}{{Astrophys. J. Lett.}}

\newcommand{\jgr}{{J. Geophys. Res.}}
\newcommand{\solphys}{{Solar Phys.}}
\begin{document}

\title{Relativistic magnetic reconnection in pair plasmas and its astrophysical applications\thanks{All authors contributed equally to this review.}
}

\titlerunning{Relativistic pair reconnection in astrophysics}        

\author{D.~Kagan\and L.~Sironi \and B.~Cerutti \and D.~Giannios}

\institute{
           D. Kagan \at Racah Institute of Physics, The Hebrew University of Jerusalem, Jerusalem 91904, Israel
           	\\Raymond and Beverly Sackler School of Physics and Astronomy, Tel Aviv University, Tel Aviv 69978, Israel\\
              \email{daniel.kagan@astro.huji.ac.il}
              \and
				L. Sironi \at Harvard-Smithsonian Center for Astrophysics, Cambridge, MA, 02138, USA\\ \email{lsironi@cfa.harvard.edu}
           \and
B. Cerutti \at Department of Astrophysical Sciences, Princeton University, Princeton, NJ 08544, USA \\
              \email{bcerutti@astro.princeton.edu}
	\and 
		D. Giannios \at Department of Physics and Astronomy, Purdue University, 525 Northwestern Avenue, West Lafayette, IN, 47907, USA\\ \email{dgiannio@purdue.edu}
}

\date{Received: date / Accepted: date}

\maketitle

\begin{abstract}
This review discusses the physics of magnetic reconnection -- a process in which the magnetic field topology changes and magnetic energy is converted to kinetic energy -- in pair plasmas in the relativistic regime. We focus on recent progress in the field driven by theory advances and the maturity of particle-in-cell codes. This work shows that fragmentation instabilities at the current sheet can play a critical role in setting the reconnection speed and affect the resulting particle acceleration, anisotropy, bulk flows, and radiation. Then, we discuss how this novel understanding of relativistic reconnection can be applied to high-energy astrophysical phenomena, with an emphasis on pulsars, pulsar wind nebulae, and active galactic nucleus jets.

\keywords{acceleration of particles \and galaxies: active \and instabilities \and magnetic reconnection \and pulsars: general \and radiation mechanisms: non-thermal \and relativistic processes}
\end{abstract}

\section{Introduction}\label{intro}

Magnetic reconnection is a common phenomenon in which the topology of magnetic field lines is changed and magnetic energy is converted to kinetic energy.  Interpretations of space plasma measurements \citep[e.g.,][] {2008NatPh...4...19C, 2011PhRvL.107p5007O} and astronomical observations suggest that reconnection occurs in many places in the Universe. Because the length scale of magnetic fields in astrophysical plasmas is extremely large, of order the size of astrophysical sources, while low plasma resistivity means that the characteristic scale of dissipation is very small,  magnetic field lines are typically ``frozen'' into the astrophysical plasma, inhibiting dissipation. The topological change in the field lines produced by reconnection can break flux freezing and facilitate dissipative energy conversion.

In this review, we focus on reconnection in pair plasmas in the relativistic regime, in which the magnetic energy before the fields reconnect is significantly greater than the total enthalpy of the particles, so that the particles become relativistic when they enter the reconnection region.   This condition is precisely stated as
\begin{equation}
\sigma\equiv \frac{B^2}{4\pi m n c^2 w_n}>1
\label{eq:sigma}
\end{equation}
where $B$ is the magnetic field, $n$ is the total particle number density including all species, and $w_n$ is the enthalpy per particle (assumed to be the same for both species), given by $w_n=\gamma +P/(m n c^2)$, where $\gamma$ is the mean particle Lorentz factor and $P$ is the particle pressure. \footnote{If $w_n\gg1$ but $\sigma<1$, the plasma is initially relativistic but reconnection is typically weak, so the relativistic reconnection discussed in this review typically fulfils condition (\ref{eq:sigma}).} 

Relativistic reconnection may be of importance in astrophysical magnetically dominated systems such as Pulsar Wind Nebulae (PWN), as well as relativistic jets in Active Galactic Nuclei (AGN) or Gamma Ray Bursts (GRB) which may be magnetically dominated. The observed radiation from such systems is typically highly energetic and nonthermal. Because shock acceleration of particles through the Fermi process is likely to be inefficient in magnetically dominated flows \citep{2011ApJ...726...75S,sironi_13}, it is expected that reconnection is responsible for the acceleration of high energy particles and the production of radiation in these magnetically dominated systems. The role of relativistic reconnection in particle acceleration and radiation is a primary subject of this review.

This paper is organised as follows. In the remainder of Section \ref{intro}, we review simple models of relativistic reconnection (Section \ref{models}) and discuss  the physics of particle acceleration in relativistic reconnection (Section \ref{intro_rad}) In Section \ref{simulations}, we discuss simulations of relativistic reconnection and the resulting particle acceleration, anisotropies, and bulk flows. In Section \ref{applications}, we explore the application of relativistic reconnection in astrophysical systems; this section includes predictions of the radiation spectrum resulting from reconnection in those systems. Finally, in Section \ref{conclusions} we present our conclusions.

\subsection{Models of Reconnection}\label{models}

We now discuss models of reconnection in detail. Whenever regions of opposite magnetic polarity are present, Maxwell's equations imply that there will be a current sheet in between. In this current layer, magnetic field lines can diffuse across the plasma to reconnect at one or more X-lines. During reconnection, magnetized plasma approaches the central plane of the current layer with an asymptotic inflow velocity $v_{\rm in}$, which is also known as the reconnection velocity. After passing the X-line, plasma is expelled from the vicinity of the X-line to either side at the outflow velocity $v_{\rm out}$, which is typically assumed to equal the characteristic speed of magnetic disturbances in plasma, the Alfv\'en velocity $v_{\rm A}$. In the relativistic regime, $v_{\rm A} =c\sqrt{\sigma/(1+\sigma)}\sim c$. The dimensionless reconnection rate is usually defined as $r_{\rm rec}\equiv v_{\rm in}/v_{\rm out}$.
 
 Outside of the current sheet, non-ideal effects are negligible and the magnetohydrodynamic (MHD) condition 
\begin{equation}
\mathbf{E} +\frac{1}{c} \langle \mathbf{v} \rangle \times \mathbf{B}=0, \label{eq:mhd}
\end{equation}
holds, where $\mathbf{E}$  is the electric field, $\mathbf{B}$ is the magnetic field, and  $\langle\mathbf{v}\rangle$ is the mean particle velocity. 

In a steady-state configuration which is quasi-two dimensional and does not vary strongly perpendicular to the plane of reconnection, the electric field throughout the
reconnection region $\mathbf{E}_{\rm rec}$ is uniform and can be found by applying the condition (\ref{eq:mhd}) outside the current sheet, giving
\begin{equation}
{\mathbf E}_{\rm rec}=-\frac{1}{c} ({\mathbf v}_{\rm in}\times {\mathbf B_0}),
\label{eq:estructure}
\end{equation}
where ${\mathbf B_0}$ is the reversing magnetic field outside the current sheet. Because there is no velocity flow inside the current sheet, the electric field there is sustained by some non-ideal effect which is responsible for dissipation. The reconnection rate $r_{\rm rec}$ may be related to the electric field by the equation
\begin{equation}
\label{recE}
r_{\rm rec}\equiv \frac{v_{\rm in}}{v_{\rm out}}=\frac{E_{\rm rec}}{(v_{\rm A}/c)B_0} .
\end{equation}

\subsubsection{Sweet-Parker resistive and kinetic  relativistic reconnection}

Defining $\delta$ and $L$ to be the thickness and length of the current sheet, the conservation of mass from the reconnection inflow to the outflow in an incompressible plasma requires
\begin{equation}
\frac{\delta}{L}=r_{\rm rec}=\frac{v_{\rm in}}{v_{\rm out}} \sim \frac{v_{\rm in}}{v_{\rm A}} .
\label{sprec}
\end{equation}

This equation is not always applicable to relativistic reconnection due to the possible presence of relativistic bulk flows which violate the incompressibility assumption, but it does apply in the simple steady-state models we discuss in this section.

In the Sweet-Parker resistive model of reconnection, $L$ is taken to be the macroscopic length scale of the magnetic field, while the thickness $\delta$ is determined by the dissipation rate that can be sustained by resistivity. The dimensionless parameter that determines the importance of collisional resistivity is the Lundquist number $S\equiv v_{\rm A} L/\eta$, where $\eta$ is the magnetic diffusivity produced by resistivity.  \citet{2005MNRAS.358..113L} has shown that the reconnection rate for relativistic Sweet-Parker resistive reconnection is 
 
\begin{equation}
r_{\rm rec}=\frac{\delta}{L}\sim \frac{1}{\sqrt{S}},
\end{equation} 

which is identical to the result for non-relativistic Sweet-Parker resistive reconnection. Since the Lundquist number $S$ is very large in astrophysical plasmas (depending on the application, $S\sim 10^{20}$ may be a typical value), Sweet-Parker reconnection is extremely slow. On the other hand, solar flares are believed to be powered by magnetic reconnection requiring that $v_{\rm in}/v_{\rm A}\sim 0.1$! 

Since the collisional resistivity is often extremely small in magnetically dominated astrophysical plasmas, kinetic effects resulting from individual particle motions are likely to be more important than resistivity in many systems.

The characteristic frequency of kinetic effects is the plasma oscillation frequency $\omega_{\rm p}$, given by 
\begin{equation}
\omega_{\rm p}=\sqrt{\frac{4 \pi n q^2 }{w_n m}},
\label{plasma}
\end{equation} 
where $q$ is the charge of the particles.  Kinetic effects become important on spatial scales smaller than the corresponding inertial length $c/\omega_{\rm p}$ (also known as ``skin depth''). \citet{2014PhRvL.113d5001C} have shown that when kinetic effects are important, the reconnection rate in the relativistic case is given by 
\begin{equation}
r_{\rm rec}=\frac{c}{\omega_{\rm p}L}.
\end{equation} 

Because $c/ \omega_{\rm p}$ is small compared to the macroscopic scale $L$ of the field lines, steady-state Sweet-Parker kinetic reconnection is still relatively slow.

\subsubsection{Fast reconnection and the tearing and plasmod instabilities}

There have been many attempts to identify effects that would result in current sheets with smaller aspect ratios $L/\delta$, to allow for faster reconnection. The most basic of these models is the Petschek mechanism \citep{1964NASSP..50..425P}, which assumes that oblique slow shocks are present around a central X-point, and they effectively limit the length of the reconnection region. Simulations in the non-relativistic regime have found that this configuration is unstable unless an anomalous localised resistivity is present in the center of the reconnection layer, i.e., at the X-line \citep{2000PhPl....7.4018U}. If the aspect ratio of the reconnection region is larger than $\sim100$, oblique slow shocks can form at the end of the reconnection exhausts, \citep{2012PhPl...19b2110L,2012JGRA..117.1220H}, but it is uncertain whether these shocks are analogous to those in the Petschek model.  Despite the difficulty in confirming the viability of this mechanism, the name ``Petschek reconnection'' is often used to describe fast reconnection because kinetic effects can produce an effective anomalous resistivity. Below, we occasionally use the relativistic ``Petschek'' model derived by \citet{2005MNRAS.358..113L} to parameterise the properties of fast reconnection in the relativistic regime.

Most other models of fast reconnection focus on the effects of instabilities in the current layer. In any current sheet, the oppositely oriented fields constitute a source of free energy. An important instability that draws on this energy is the tearing instability, which at the same time mediates and is mediated by reconnection. The tearing instability produces an alternating series of narrow X-lines where reconnection can occur, separated by large flux ropes. In turn, steady reconnection equilibria contain thin current sheets, which themselves can be unstable to the tearing instability. The nonideal effect that violates flux freezing to produce reconnection at these X-lines may be collisional resistivity, or it may arise from kinetic effects, so the tearing instability, like reconnection, can take both resistive and kinetic forms. The growth rate of the tearing instability depends strongly on the width of the current sheet. For fast growth, the sheet width must be comparable to those associated with resistive or kinetic reconnection \citep{2000mrp..book.....B, 2007PPCF...49.1885P}.

A Sweet-Parker resistive current sheet is thin enough that a resistive instability of the Sweet-Parker current sheet, called the plasmodia instability, may break the sheet into X-lines and magnetic islands, thus lowering its aspect ratio $L/\delta$ and leading to relatively fast reconnection rates $r_{\rm rec}\sim 0.01$ even at high Lundquist numbers, for which the unperturbed Sweet-Parker reconnection would be extremely slow \citep{2007PhPl...14j0703L, 2009PhRvL.103j5004S, 2010PhPl...17f2104H}. However, the corresponding reconnection rate in the relativistic case is significantly lower,  $r_{\rm rec}\sim 0.0001$ \citep{2011MNRAS.418.1004Z}. In a long kinetic current sheet whose width is comparable to the skin depth, the kinetic tearing instability can grow quickly and break up the current sheet into X-lines and flux ropes, which can result in fast reconnection at $r_{\rm rec}\sim 0.1$ \citep[e.g.,] []{2001JGR...106.3737B}. A phase diagram of reconnection has been proposed uniting Sweet-Parker and plasmoid configurations for resistive and kinetic reconnection, with the transition from resistive to kinetic reconnection occurring when the Sweet-Parker sheet width approaches the skin depth, and the transition from Sweet-Parker to plasmoid configurations occurring as the aspect ratio of the reconnection region increases \citep{2011PhPl...18k1207J, 2014PhRvL.113d5001C}. The transition between resistive and kinetic regimes has been proposed as a possible explanation of observed variability in reconnection sites \citep{2008ApJ...688..555G} and the onset of fast reconnection far from the central engine in a Poynting flux model of GRBs \citep{2012MNRAS.419..573M}.  In this review, we focus on the study of kinetic relativistic reconnection and the particle acceleration and radiation that can be produced by such reconnection, because kinetic effects will often dominate in relativistic magnetically dominated astrophysical plasmas, which are typically nearly collisionless.

\subsection{Particle acceleration and radiation in reconnection}\label{intro_rad}

As discussed earlier, it is thought that magnetic reconnection is likely to be responsible for the acceleration of particles in systems that are magnetically dominated. As particles cross the current sheet at the X-line, they are forced to return into the current sheet by the reversing magnetic field, following Speiser orbits \citep{1965JGR....70.4219S}. Particles following such orbits can be accelerated in the direction perpendicular to the plane of reconnection \citep[e.g. ][]{2001ApJ...562L..63Z} by the reconnection electric field. Other acceleration mechanisms, in both X-lines and flux ropes, have been found in kinetic simulations for a review of these mechanisms see \citet{2010ApJ...714..915O}, as well as the discussion in Section \ref{acceleration}. 

The energy gain per unit time for a charged particle accelerated electromagnetically is given in general by

\begin{equation}
\frac{dW}{dt} =q \mathbf{E} \cdot \mathbf{v}\sim qEc,
\label{eq:accel}
\end{equation}  

Particles accelerated in relativistic magnetically dominated systems are typically thought to radiate via the synchrotron mechanism, which tends to place a fundamental constraint on the maximum energy of electromagnetically accelerated particles.
The total synchrotron power emitted by a particle is given approximately by
\begin{equation}
 \frac{dW}{dt}\sim \frac{2q^4B^2\gamma^2}{3m^2c^3}. 
\label{eq:synch}
\end{equation}  
In regions where the MHD condition (\ref{eq:mhd}) holds, $E\le B$ and setting $E=B$ allows the derivation of a maximum $\gamma$ for charged particles, which corresponds to a maximum radiation frequency referred to as the synchrotron burnoff limit. However, during reconnection particles experiencing extreme acceleration at the X-line can spend most of their time deep in the reconnection layer where  $E>B$ \citep{2011ApJ...737L..40U}. Thus, they are able to evade this restriction and produce radiation beyond the burnoff limit, as we demonstrate in Section \ref{PWN}.

\section{Particle-in-cell simulations of reconnection}\label{simulations}

\subsection{Numerical setup}\label{setup}

\subsubsection{Numerical techniques}

The most common method for simulating the kinetic dynamics of a reconnecting plasma involves the use of a particle-in-cell (PIC) code that evolves the discretized  equations of electrodynamics -- Maxwell's equations and the Lorentz force law. See \citet{1991ppcs.book.....B} for a detailed discussion of this method.  PIC codes can model astrophysical plasmas from first principles, as a collection of charged macro-particles that are moved by integration of the Lorentz force. Each macroparticle represents many physical particles. Currents associated with the macro-particles are deposited on a grid on which Maxwell's equations is discretized. Electromagnetic fields are then advanced via Maxwell's equations, with particle currents as the source term. Finally, the updated fields are extrapolated to the particle locations and used for the computation of the Lorentz force, so the loop is closed self-consistently. So long as current deposition is the only effect of the macro-particles on the field quantities, charge conservation is ensured. This approach is capable of treating all effects present in collisionless plasmas, including particle acceleration to high energies.  To ensure that kinetic effects are resolved in the simulation, it is necessary that the grid spacing be much smaller than the skin depth $c/\omega_{\rm p}$, and that the timestep be much smaller than the corresponding timescale $\omega_{\rm p}^{-1}$. To ensure that the momentum space distribution is adequately sampled, keep particle noise at a low level, and reduce the effects of unphysical collisions due to the relatively small number of particles in a Debye sphere, it is necessary that there be several particles per cell for each particle species.

\subsubsection{The Harris current sheet}

The starting equilibrium of most reconnection simulations is the Harris current sheet, which is an exact 1D equilibrium of plasma physics \citep{harris62}. It is characterised by the field profile
\begin{equation}
{\mathbf B}=B_0 \tanh \frac{y}{\delta}\ \hat{{\mathbf x}} +\kappa B_0 \hat{{\mathbf z}},
\label{eq:harris_field}
\end{equation} 

where $\delta$ is the half-thickness of the current sheet, which must be of the same order as $c/\omega_{\rm p}$ for fast reconnection to occur. The quantity $\kappa$ sets the relative strength of a uniform ``guide'' field (orthogonal to the reconnection plane) which may be present in realistic reconnection configurations. For most of the discussion below, we will assume $\kappa=0$ for the sake of simplicity.

 The particles within the current sheet in the Harris equilibrium are initialised in a drifting Maxwell-Juttner thermal distribution in which positively and negatively charged particles have equal and opposite bulk velocities $\boldsymbol{\beta}_+=-\boldsymbol{\beta}_-=\boldsymbol{\beta}$ (in units of the speed of light) and drifting Lorentz factors of $\gamma_d=1/\sqrt{1-\boldsymbol{\beta}^2}$.  

The density profile of the Harris current sheet including both electrons and positrons in the simulation frame is

\begin{equation}
\label{eq:density_profile}
n=n_0 \ {\rm sech}^2\ \frac{y}{\delta},
\end{equation}

Pressure equilibrium requires that $B_0^2=8 \pi n_0 T_0$, where $T_0$ is the temperature of the particles (in units of $m c^2$) in the current sheet including the Boltzmann constant in the simulation frame. Amp\`ere's Law requires that
\begin{equation}
\boldsymbol{\beta}_+=-\boldsymbol{\beta}_-=B_0 /(4\pi n_0 q \delta) (-\hat{{\mathbf z}}).
\end{equation}

This simple configuration is unstable to the tearing instability and is useful for studying reconnection. An additional uniform background population of particles with rest-frame density $n_b$ and no drift velocity is typically added to the current sheet population. Thus, the total density in the simulation frame of all particles in the middle of the current sheet is $n_0 +n_{\rm b}$, whereas the total density in the background plasma away from the current sheet is $n_{\rm b}$. Using the expression for pressure equilibrium above allows us to express the value of $\sigma$ far from the current sheet as
\begin{equation}
\sigma=\frac{2 n_0 T_0 }{n_b w_{n,\rm b}}.
\end{equation}
where $w_{n,b}$ is the mean enthalpy of the particles in the background plasma. Note that the value of $n_0 T_0$ is a Lorentz invariant. This equilibrium can be modified while retaining the same value of $\sigma_b$ by increasing the temperature $T_0$ and decreasing the value of $n_0/n_b$ to produce an equilibrium with less density contrast but a difference in temperature between the populations; for a detailed discussion, see \citet{2013A&A...558A.133M}. This modification is used in the simulations in this paper.

While the Harris sheet is the most common initial condition for studying reconnection, it should be mentioned that there are other possibilities. Reconnection can be initialised using a force-free current sheet \citep{guo14}, and dynamical scenarios such as X-point collapse \citep{2014PhPl...21a2901G}. Finally, fully three dimensional configurations \citep[][ and references therein]{2011AdSpR..47.1508P}
are likely to be the most realistic starting points for simulation, but only a few PIC simulations have used such configurations \citep{2012ApJ...759L...9B,PhysRevLett.111.045002}. 
\begin{figure*}[!tbp]
\begin{center}
\includegraphics[width=1.05\textwidth]{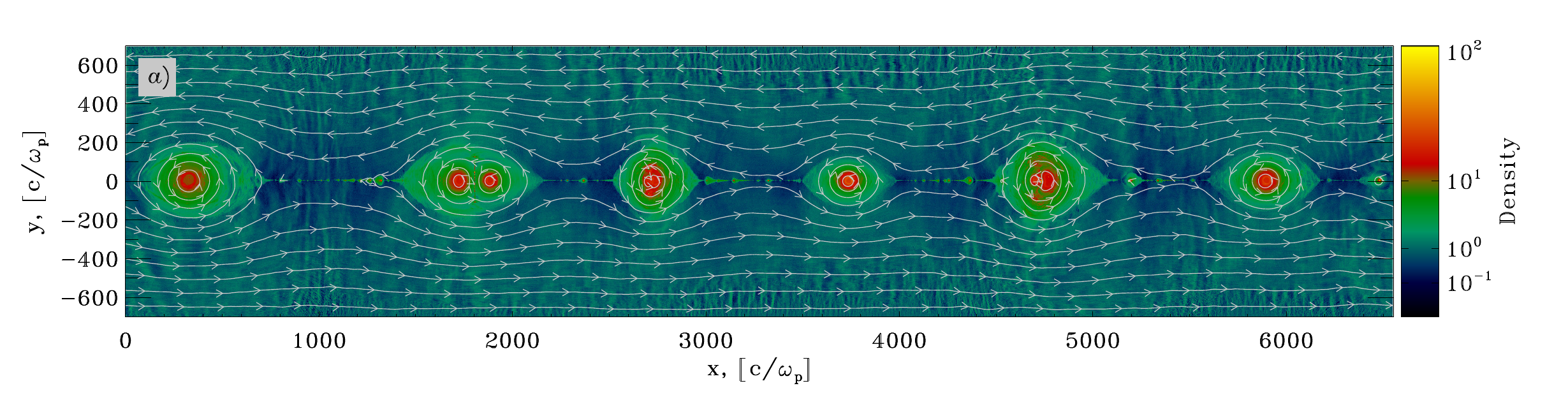}
\caption{Structure of the particle density in the reconnection layer at $\ompt=3000$, from a 2D simulation of $\sigma=10$ reconnection  presented in \citet{2014ApJ...783L..21S}.}
\label{fig:fluid2da}
\end{center}
\end{figure*}

\begin{figure*}[!tbp]
\begin{center}
\includegraphics[width=1.05\textwidth]{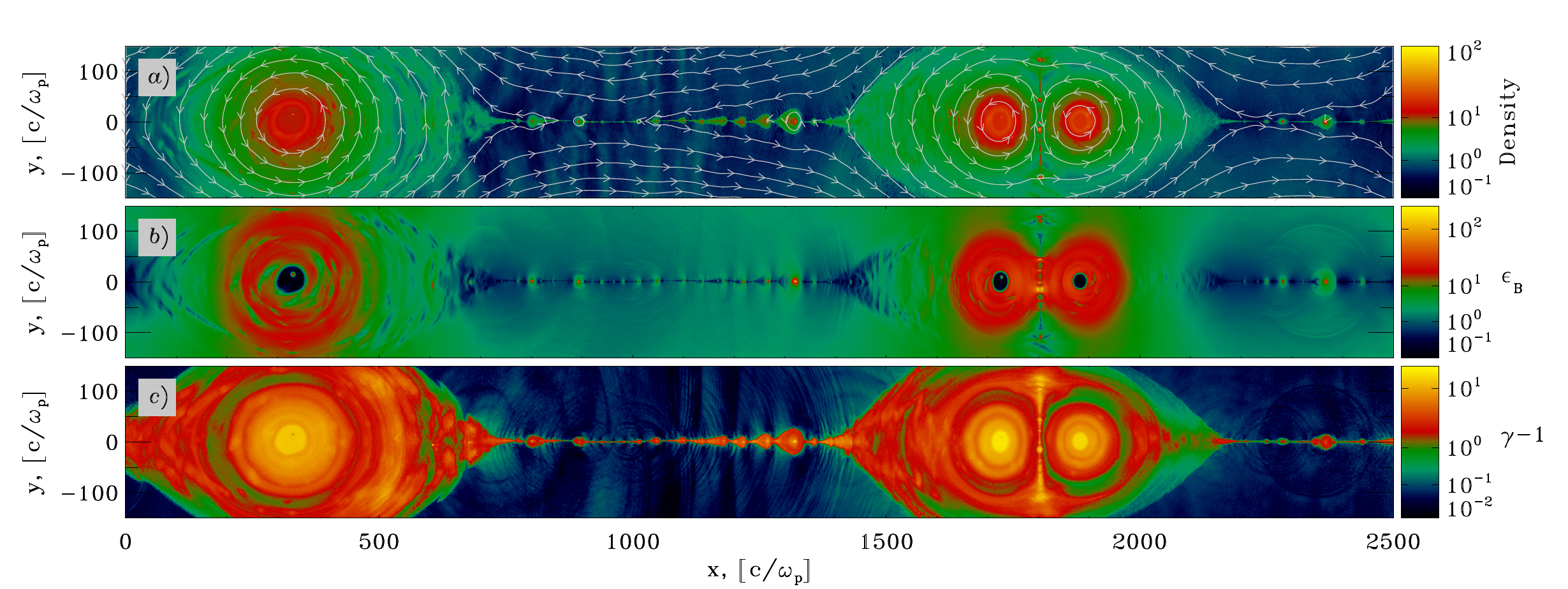}
\caption{Structure of the reconnection layer at $\ompt=3000$, from a 2D simulation of $\sigma=10$ reconnection discussed in \citet{2014ApJ...783L..21S}. This figure is a zoom-in at $0\leq x\leq 2500\comp$ of \fig{fluid2da}. We present (a) particle density, in units of the number density far from the current sheet (with overplotted magnetic field lines), (b) magnetic energy fraction $\epsilon_B=B^2/8\pi m n_{\rm b} c^2$ and (c) mean kinetic energy per particle.}
\label{fig:fluid2d}
\end{center}
\end{figure*}

\begin{figure}[!tbp]
\begin{center}
\includegraphics[width=0.7\textwidth]{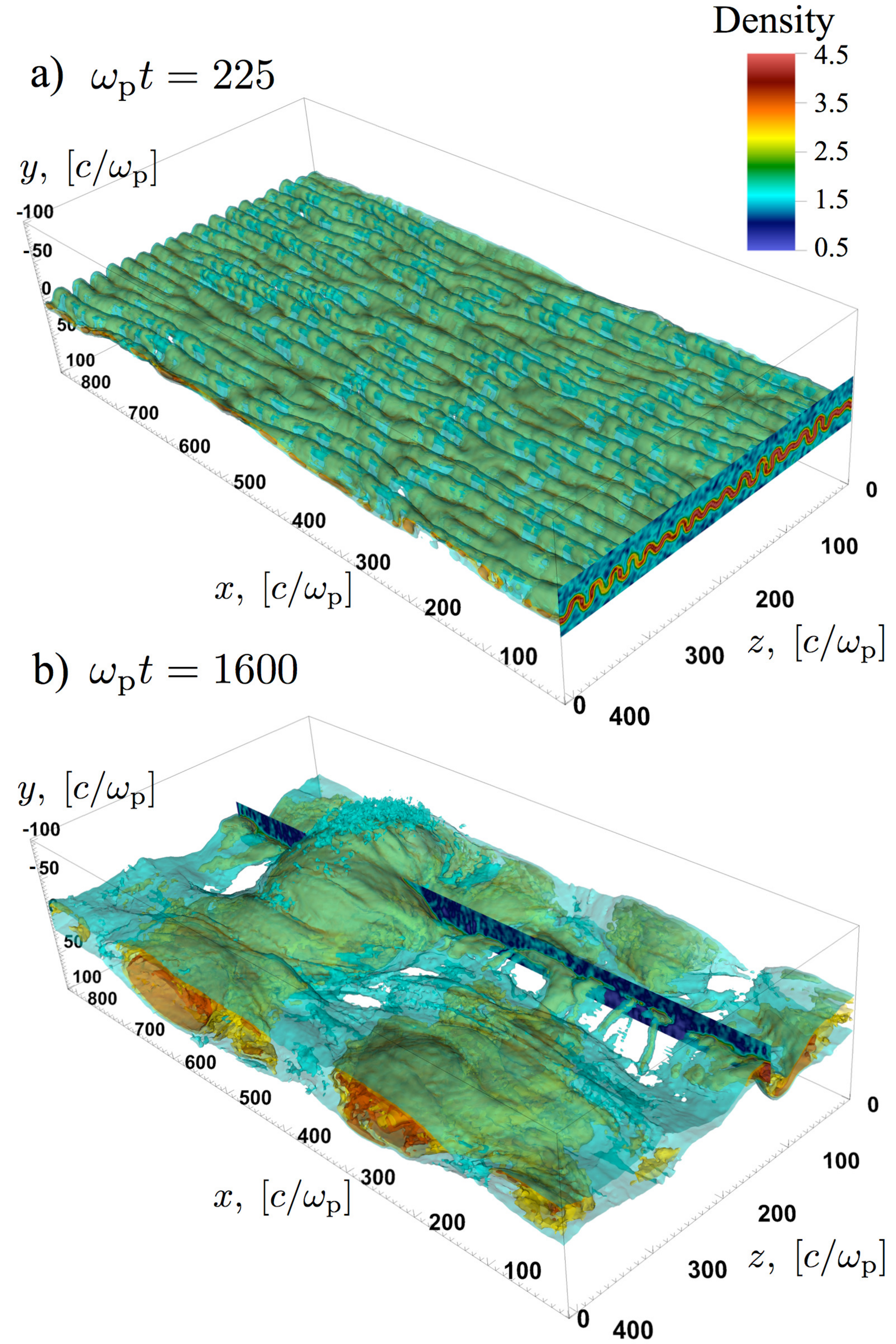}
\caption{Structure of the particle density at two different times: (a) $\ompt=250$ and (b) $\ompt=1600$. The plot refers to a 3D simulation of $\sigma=10$ reconnection without a guide field,  presented in \citet{2014ApJ...783L..21S}. The 2D slices in the top and bottom panels (at $x=0$ and $z=130\comp$, respectively) show the particle number density in that plane.}
\label{fig:fluid3d}
\end{center}
\end{figure}

\subsection{Structure of the reconnection layer}\label{structure}
We now present the structure and the  dynamics of the reconnection layer, discussing the results of 2D and 3D PIC simulations. We concentrate on the case of an electron-positron plasma, which has been most widely explored in the literature, both in 2D  \citep[][]{2001ApJ...562L..63Z,zenitani_hoshino_05b,zenitani_07,zenitani_hesse_08b,jaroschek_04,jaroschek_08b,bessho_05,bessho_07,bessho_12,daughton_07,lyubarsky_liverts_08, 2012ApJ...754L..33C, 2013ApJ...770..147C, 2014arXiv1409.8262W} and in 3D  \citep[][]{zenitani_08,yin_08,liu_11,2011ApJ...741...39S, sironi_spitkovsky_12,kagan_13,2014ApJ...782..104C,2014ApJ...783L..21S,guo14}. 
The physics of relativistic electron-proton  reconnection, yet still at an early stage of investigation, shows remarkable similarities with electron-positron reconnection \citep{melzani14}.

 As described above, the reconnection layer is set up in Harris equilibrium, with the magnetic field reversing at $y=0$. For the sake of simplicity, we discuss here the case of anti-parallel fields, without a guide field component.
The strength of the alternating fields is parameterized by the magnetization $\sigma$ defined in Eq. \ref{eq:sigma}. Here, we assume that the background plasma far from the current sheet is cold, so $w_n\sim1$ and $\sigma=B_0^2/4\pi m n_{\rm b} c^2$. 

As a result of the tearing instability, the reconnection layer fragments into a series of magnetic islands (or flux tubes), separated by X-points. Over time, the islands coalesce and grow to larger scales (\citealt{daughton_07} have described a similar evolution in non-relativistic reconnection). The structure of the reconnection region at late times is presented in \fig{fluid2da}, from a large-scale 2D simulation in a $\sigma=10$ pair plasma presented in \citet{2014ApJ...783L..21S}.  By zooming into the region $0\lesssim x\lesssim 2500\comp$ (here, the inertial length $\comp$ is measured taking the density far from the current sheet), we see that each X-line is further fragmented into a number of smaller islands. This is a result of the secondary tearing mode (or ``plasmoid instability'') discussed by \citet{2010PhRvL.105w5002U}. The secondary islands lie at $700\comp\lesssim x \lesssim 1400\comp$ in \fig{fluid2d}. They are overdense (\fig{fluid2d}a), filled with hot particles (\fig{fluid2d}c) and confined by strong fields (\fig{fluid2d}b). In between each pair of secondary islands, a secondary X-point mediates the transfer of energy from the fields to the particles. As shown in the next section, efficient particle acceleration occurs at the X-points.

The reconnection rate is $r_{\rm rec}\equiv v_{\rm in}/v_{\rm out}\sim v_{\rm in }/c\simeq 0.08$ for $\sigma=10$, nearly constant at late times. The reconnection rate depends on the plasma magnetization. In the case of vanishing guide field,  \citet{2014ApJ...783L..21S} quote that the reconnection rate in 2D increases from $r_{\rm rec}\simeq 0.03$ for $\sigma=1$ to $r_{\rm rec}\simeq 0.12$ for $\sigma=30$, and it is nearly independent of $\sigma$ for larger magnetizations, in agreement with the analytical model by \citet{2005MNRAS.358..113L}. 

After entering the current sheet, the flow is advected towards the large magnetic islands by the tension force of the reconnected magnetic field (in \fig{fluid2d}a-c, the major islands are at $200\comp\lesssim x\lesssim500\comp$ and $1600\comp\lesssim x\lesssim1900\comp$). 
Pushed by the ram pressure of the reconnection outflows, the major islands move along the  layer, merging with neighboring islands. A merger event in indeed seen at $x\sim 1800\comp$ in \fig{fluid2d}. The current sheet formed between the two merging islands is unstable to the tearing mode, and it breaks into a series of secondary islands  along the $y$ direction (orthogonal to the primary current sheet). 

The evolution of 3D reconnection at late times parallels closely the 2D physics described above, even in the absence of a guide field.\footnote{The presence of a strong guide field orthogonal to the reconnecting plane guarantees that the 3D physics will resemble the 2D results, see \citet{guo14}.} As shown in \fig{fluid3d}a, the early phases of evolution are governed by the so-called drift-kink (DK) mode \citep{zenitani_08,2014ApJ...782..104C,2014ApJ...783L..21S}. The DK instability corrugates the current sheet in the $z$ direction, broadening the  layer and inhibiting the growth of the tearing mode at early times. However, at later times the evolution is controlled by the tearing instability \citep{2014ApJ...783L..21S}, that produces in the $xy$ plane a series of magnetic islands (or rather, flux tubes), in analogy to the 2D physics. The reconnection layer at late times is organized into a few major islands (see the overdense plasmoids in \fig{fluid3d}b), separated by underdense regions (transparent in \fig{fluid3d}b) where field dissipation by reconnection is most efficient. 
In short, at late times the 3D physics parallels closely the 2D evolution presented above (yet, with a smaller reconnection rate, $r_{\rm rec}\simeq0.02$ in 3D versus $r_{\rm rec}\simeq0.08$ in 2D). As discussed in the next section, this has important implications for the acceleration performance of relativistic reconnection in 3D.

\begin{figure}[tbp]
\begin{center}
\includegraphics[width=0.8\textwidth]{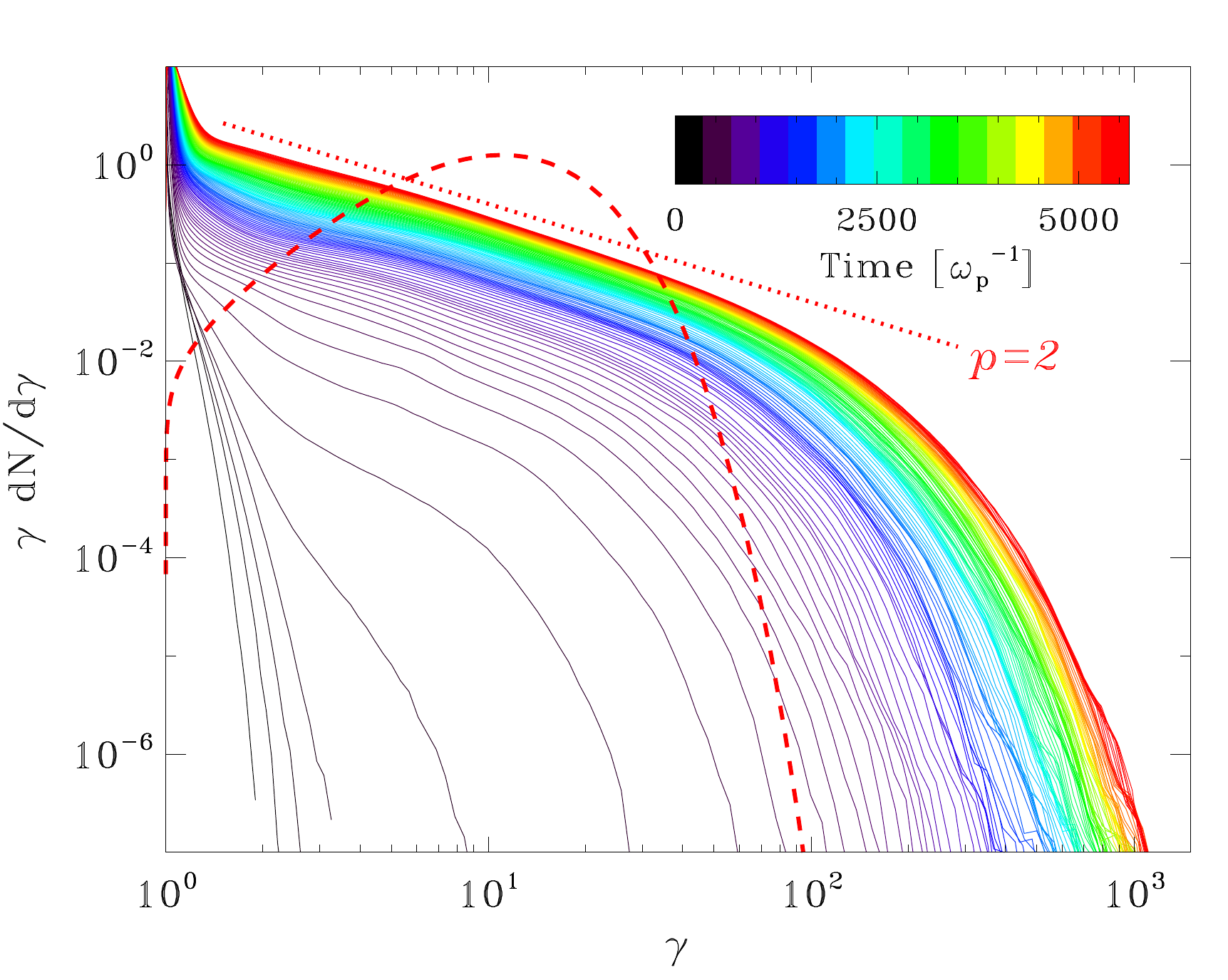}
\caption{Temporal evolution of the particle energy spectrum, from a 2D simulation of $\sigma=10$ reconnection by \citet{2014ApJ...783L..21S}. The spectrum at late times resembles a power-law with slope $p=2$ (dotted red line), and it clearly deviates from a Maxwellian with mean energy $(\sigma+1)\,mc^2$ (dashed red line, which assumes complete field dissipation).}
\label{fig:spec2d}
\end{center}
\end{figure}

\begin{figure}[tbp]
\begin{center}
\includegraphics[width=0.8\textwidth]{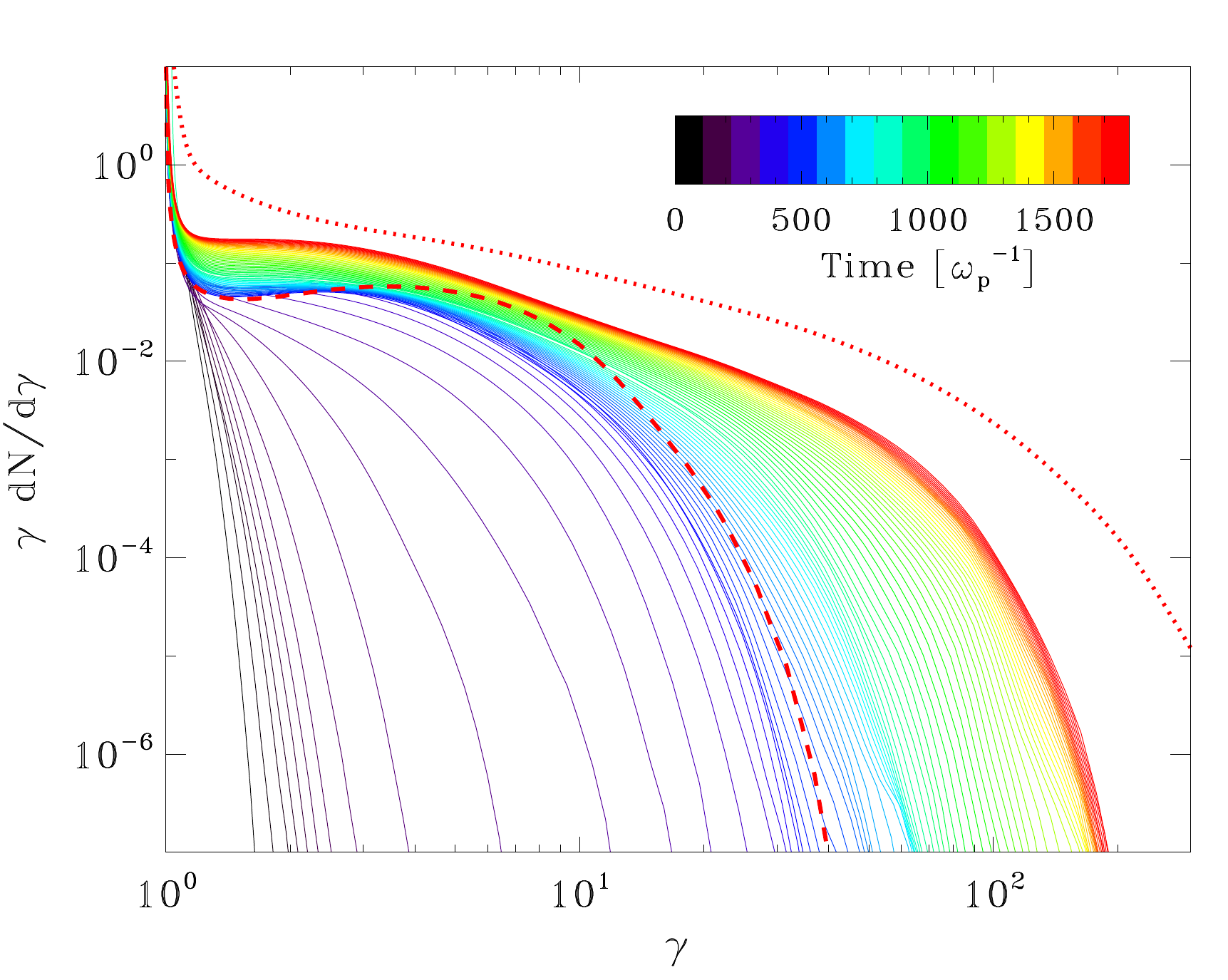}
\caption{Temporal evolution of the particle energy spectrum, from a 3D simulation of $\sigma=10$ reconnection by \citet{2014ApJ...783L..21S}. The spectra from two 2D simulations with in-plane (out-of-plane, respectively) anti-parallel fields are shown with red dotted (dashed, respectively) lines. }
\label{fig:spec3d}
\end{center}
\end{figure}

\begin{figure}[tbp]
\begin{center}
\includegraphics[width=0.8\textwidth]{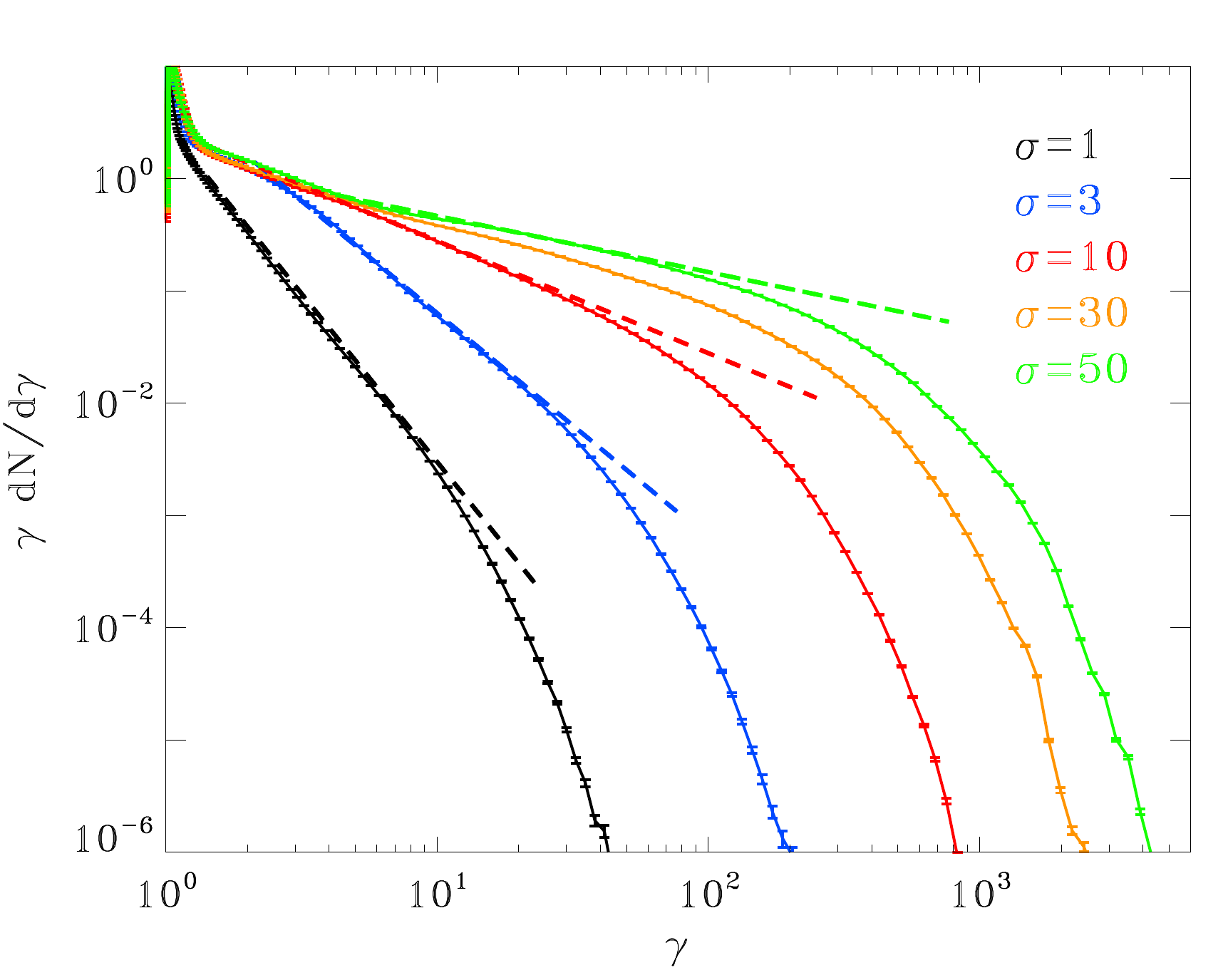}
\caption{Dependence of the spectrum on the magnetization, as indicated in the legend. The dotted lines refer to power-law slopes of $-4$, $-3$, $-2$ and $-1.5$ (from black to green).}
\label{fig:spec2db}
\end{center}
\end{figure}

\subsection{Non-thermal particle acceleration}\label{acceleration}
Relativistic reconnection is an efficient source of non-thermal particles.
In \fig{spec2d} we present the time evolution of the particle energy spectrum, from a 2D simulation of reconnection with $\sigma=10$ performed by \citet{2014ApJ...783L..21S}. A generic by-product of relativistic reconnection is the generation of a broad non-thermal spectrum extending to ultra-relativistic energies. For $\sigma=10$, the spectrum at $\gamma\gtrsim 1.5$ can be fitted with a power-law of slope $p\equiv - d\log N/d\log \gamma\sim2$ (dotted red line). The spectrum clearly departs from a Maxwellian distribution with mean energy $(\sigma+1)\,mc^2$  (red dashed line, which assumes complete field dissipation). As shown in \fig{spec2db}, the power-law slope depends on the flow magnetization, being harder for higher $\sigma$ ($p\sim1.5$ for $\sigma=50$, compare solid and dotted green lines). The slope is steeper for lower magnetizations  ($p\sim4$ for $\sigma=1$, solid and dotted black lines), approaching the result of non-relativistic reconnection, yielding poor acceleration efficiencies \citep[][]{drake_10}. In the limit $\sigma\gg1$, \citet{guo14} and \citet{2014arXiv1409.8262W} have confirmed the trend described above, arguing that the non-thermal slope asymptotes to $p\simeq 1$ for highly magnetized flows.


For magnetizations $\sigma\gtrsim10$ that yield $p\lesssim2$, the increase in maximum energy over time is expected to terminate, since the mean energy per particle cannot exceed $(\sigma+1)\,mc^2$. For a power-law of index $1<p<2$ starting from $\gamma_{\rm min}=1$, the maximum Lorentz factor should saturate at $\gamma_{\rm max}\sim[(\sigma+1)(2-p)/(p-1)]^{1/(2-p)}$. For $\sigma\lesssim 10$ (where $p\gtrsim 2$), the increase in maximum energy does not stop, but it slows down at late times.

In short, 2D simulations of relativistic reconnection produce hard populations of non-thermal particles. However, the structure of X-points in 3D is different from 2D, as emphasized in the previous section. In particular, the DK mode is expected to result in heating, not in particle acceleration \citep{zenitani_07}. \fig{spec3d} presents the temporal evolution of the particle spectrum in a 3D simulation with $\sigma=10$, by \citet{2014ApJ...783L..21S}. The spectrum at early times is quasi-thermal (black to blue lines in \fig{spec3d}), and it resembles the distribution resulting from the DK mode (the red dashed line shows the spectrum from a 2D simulation with out-of-plane anti-parallel fields, to target the contribution of the DK mode). As discussed above, the DK mode is the fastest to grow, but the sheet evolution at late times is controlled by the tearing instability, in analogy to the 2D physics with in-plane fields. In fact,  the spectrum at late times (cyan to red lines in \fig{spec3d}) presents a pronounced high-energy power-law. The power-law slope is $p\sim2.3$, close to the $p\sim2$ index of 2D simulations with in-plane fields. With respect to the 2D spectrum (dotted red line in  \fig{spec3d}), the normalization and the upper energy cutoff of the 3D spectrum are smaller, due to the lower reconnection rate ($r_{\rm rec}\simeq 0.02$ in 3D versus $r_{\rm rec}\simeq 0.08$ in 2D), so that fewer particles enter the current sheet per unit time, where they get accelerated by a weaker  electric field $E_{\rm rec}\sim r_{\rm rec}\, B_0$. 

\begin{figure}[]
\centering
\includegraphics[width=8.0cm]{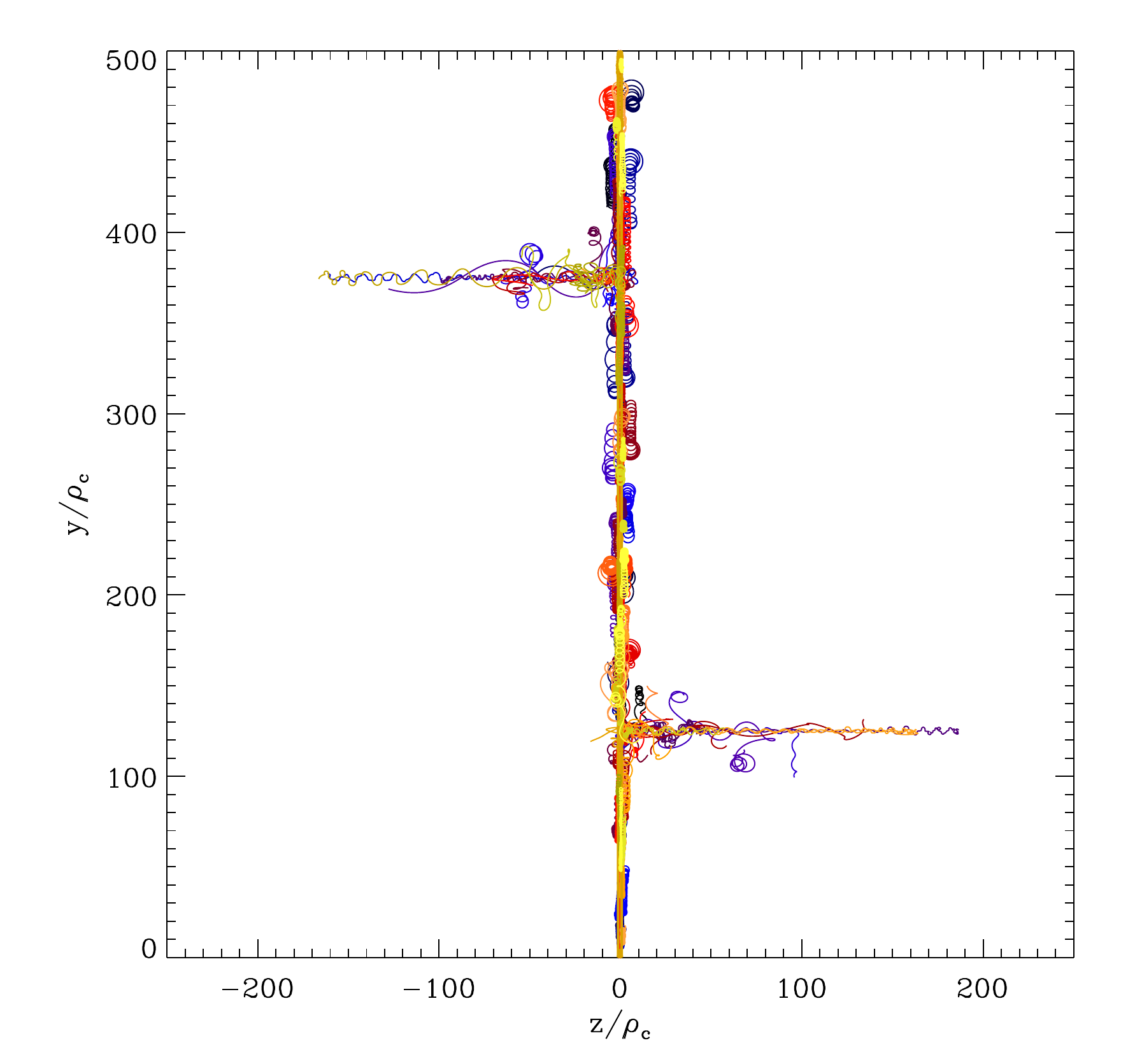}
\caption{Trajectories of a sample of 150 particles projected in the $(yz)$-plane from a 2D PIC simulation of relativistic reconnection with $\sigma=10$, and without guide field.  Each orbit are drawn with a different color to increase the readability of this figure.The simulation starts with two anti-parallel Harris sheets of temperature $kT=mc^2$ located at $y/\rho_{\rm c}\sim125$ and $375$, where $\rho_{\rm c}=mc^2/eB_0$. Particles are accelerated along the $z$-axis within the current layers where the electric field is maximum, and they follow special orbits known as relativistic Speiser orbits. The further the particle gets along the $z$-axis, the more energetic the particle will become.}
\label{fig_orbits}
\end{figure}

The mechanism of particle acceleration at X-points has been the subject of various investigations, with analytical \citep{larrabee_03,bessho_12} or numerical methods.\footnote{{ Particle acceleration in magnetic islands (as opposed to X-lines or X-points) is also widely discussed in the literature, both in non-relativistic reconnection \citep[e.g.,][]{drake_06,2010ApJ...714..915O} --- where the particles are adiabatic, and they bounce several times between the two edges of an island --- and relativistic reconnection \citep{liu_11,guo14}, where the energy gain might come just from a single bounce. However, the inflowing particles interact at first with the X-points, where they get energy from the dissipating fields. It is this first acceleration episode (that we describe below) which will establish the spectral slope and strongly affect the future history of the inflowing particles. In fact, particles accelerated to high energies at the X-point are likely to experience further acceleration via reflection off of moving magnetic disturbances (e.g., in contracting islands or in between two merging islands), which might eventually dominate the overall energy gain.}} Using test particle simulations in prescribed electromagnetic fields, \citet{2003PhPl...10..835N, 2011ApJ...737L..40U, 2012ApJ...746..148C} found that reconnection naturally produces beams of high-energy particles aligned with the reconnection electric field present within the current layer. These particles follow relativistic Speiser orbits as they are moving back and forth across the reconnection layer. For a steady Sweet-Parker configuration, \citet{2011ApJ...737L..40U} showed that the meandering width of the Speiser orbit decreases as the energy of the particle increases, i.e., the most energetic particles, {with larger Lorentz factor}, are also the most focused along the electric field (see also \citealt{2004PhRvL..92r1101K, 2007A&A...472..219C}). The properties of these special orbits are also well captured by PIC simulations \citep{2012ApJ...746..148C, 2013ApJ...770..147C}. Fig.~\ref{fig_orbits} shows the trajectory of a sample of 150 particles chosen randomly in a 2D PIC simulation with $\sigma=10$. The particle orbits are projected in the plane perpendicular to the reconnecting field, i.e., here in the $(yz)$-plane (reconnection happens in the $xy$-plane). Away from the two layers (located at $y/\rho_{\rm c}\sim125$ and $375$, with $\rho_{\rm c}=mc^2/eB_0$), the particles are well magnetized: they gyrate along the field lines and remain at $z=0$. In contrast, the particles that enter the layer are efficiently boosted along the direction of the electric field (the $z$-axis) and follow relativistic Speiser orbits. The further the particle gets along the $z$-direction, the more energetic it will be.

\begin{figure}[tbp]
\begin{center}
\includegraphics[width=0.8\textwidth]{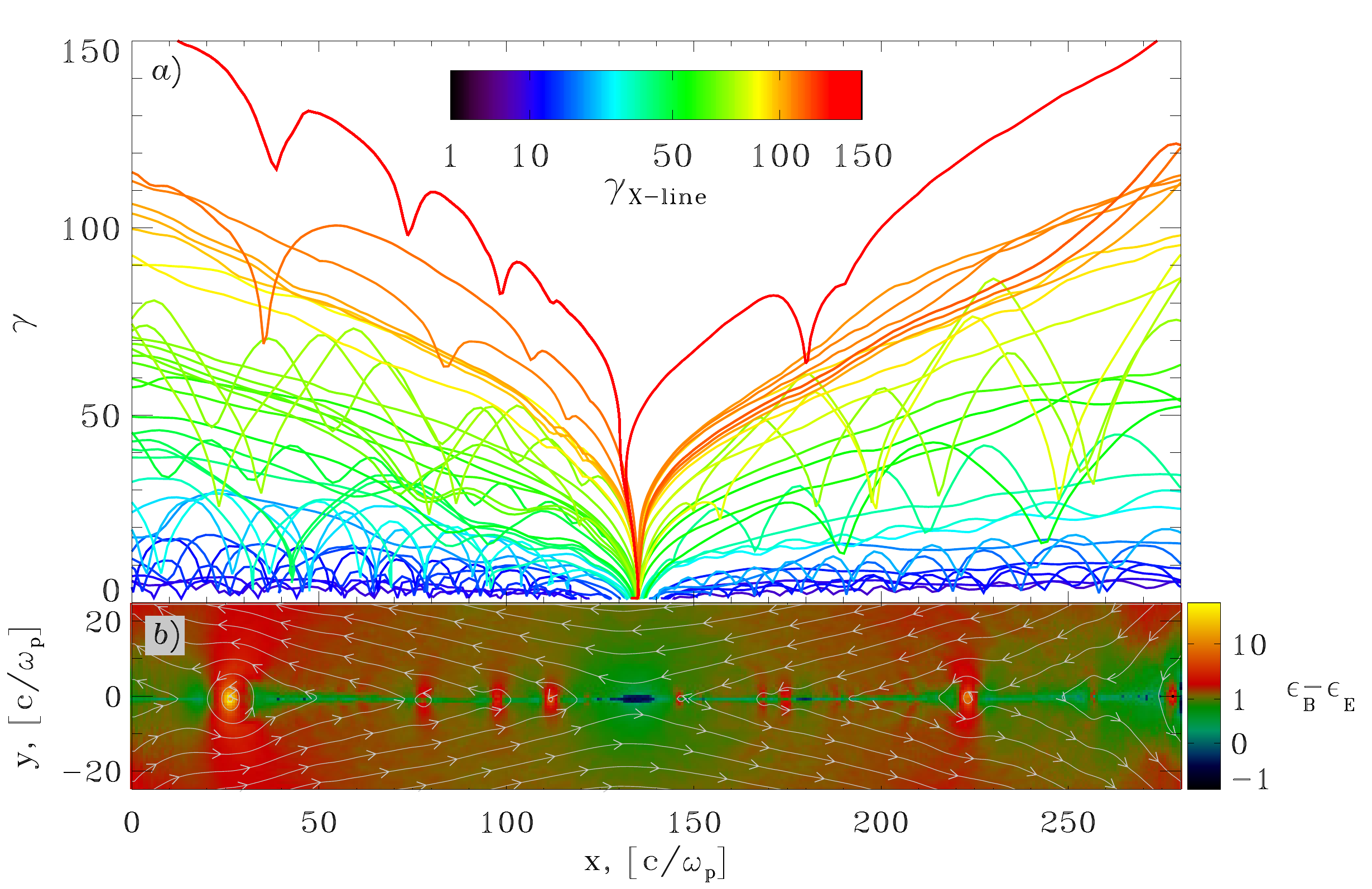}
\caption{(a) Energy evolution of a sample of selected particles interacting with a major X-point, as a function of the location $x$ along the current sheet. Colors are scaled with $\gamma_{\rm X\text{-}line}$, the  Lorentz factor attained at the outflow boundary of the X-line (at $x=0$ or $280\comp$, depending on the particle). (b) $\epsilon_B-\epsilon_E$ at the time when the particles interact with the X-point (here, $\epsilon_E=E^2/8\pi m n_{\rm b} c^2$ is the electric energy fraction).}
\label{fig:accel}
\end{center}
\end{figure}

The trajectories of a sample of particles extracted from a 2D simulation with $\sigma=10$ (in \fig{accel}, from \citet{2014ApJ...783L..21S}) also illustrate the mechanism for the formation of the power-law tail in the particle spectrum. At the X-point located at $x\sim 135\comp$ the magnetic energy is smaller than the electric energy (blue region in \fig{accel}b), so the  particles become unmagnetized and they get accelerated along $z$ by the reconnection electric field. The final energy of the particles -- the color in \fig{accel}a indicates the Lorentz factor measured at the outflow boundary of the X-line -- directly correlates with the location at the moment of interaction with the current sheet \citep[as argued in the analytical models by][]{larrabee_03,bessho_12}. Particles interacting closer to the center of the X-point (darkest blue in \fig{accel}b) are less prone to be advected away along $x$ by the reconnected magnetic field, so they can stay longer in the acceleration region and reach higher Lorentz factors (orange and red lines in \fig{accel}a).  In other words, energetic particles turn slowly into the reconnected field ($B_y$ in \fig{accel}), because the Larmor radius is proportional to $\gamma$, so that they spend even more time at the X-point than particles with lower energies. This is an argument originally proposed by \citet{2001ApJ...562L..63Z}, that may also explain the power-law nature of the spectrum (along with the impact parameter of the particles in the current sheet). Indeed, a broad power-law distribution is then established, as a result of the different energy histories of particles interacting at different distances from the X-point. 

We point out that the most energetic particles (red and orange curves in \fig{accel}) are slowly turning around the reconnected magnetic field $B_y$, and still have a positive $q\, \bmath{E}\cdot \bmath{v}$, so that they gain energy even outside of the blue region (where $|\bmath{E}|>|\bmath{B}|$). On the other hand, the green and blue particles experience also the electric fields surrounding the secondary islands, which explains the oscillations in their energy curves.

\subsection{Particle anisotropy and bulk motions}\label{anisotropy}

It is now well established that relativistic reconnection is an efficient source of non-thermal particle acceleration (see previous section). In usual astrophysical environments, these energetic particles would emit non-thermal radiation via, e.g., synchrotron or inverse Compton scattering. Due to relativistic aberrations, the radiation emitted by highly relativistic particles (with $\gamma\gg 1$) is beamed within a cone of semi-aperture angle $\sim 1/\gamma\ll 1$ along the direction of motion of the emitting particle. As a result, any anisotropy in the particle distribution results in an anisotropic distribution of radiation which is of critical importance in astronomy because the observer probes only one direction at a time. The overall energetic budget or even the shape of the particle spectrum inferred from observations could differ significantly from the isotropically averaged quantities.

\begin{figure}[]
\centering
\includegraphics[width=10cm]{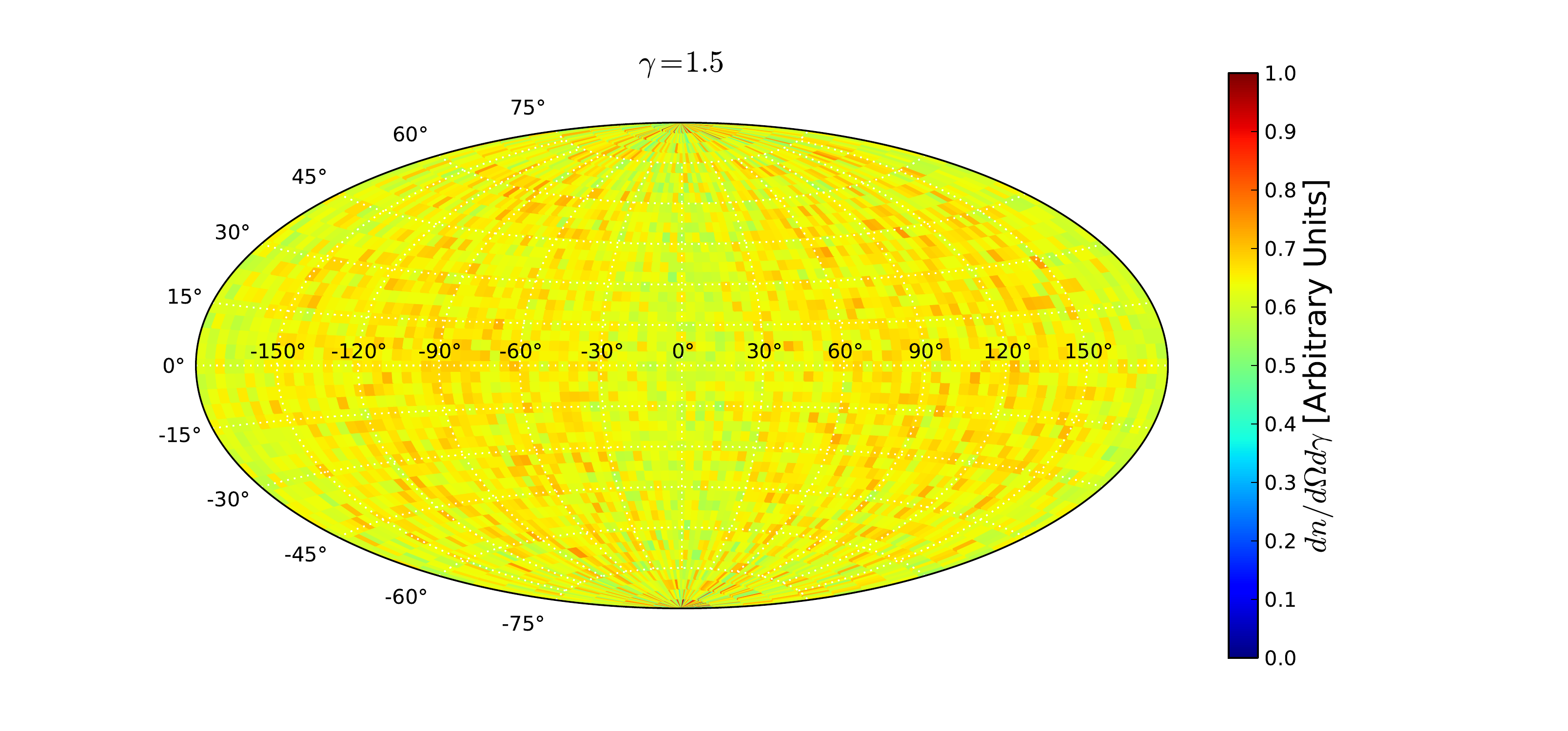}
\includegraphics[width=10cm]{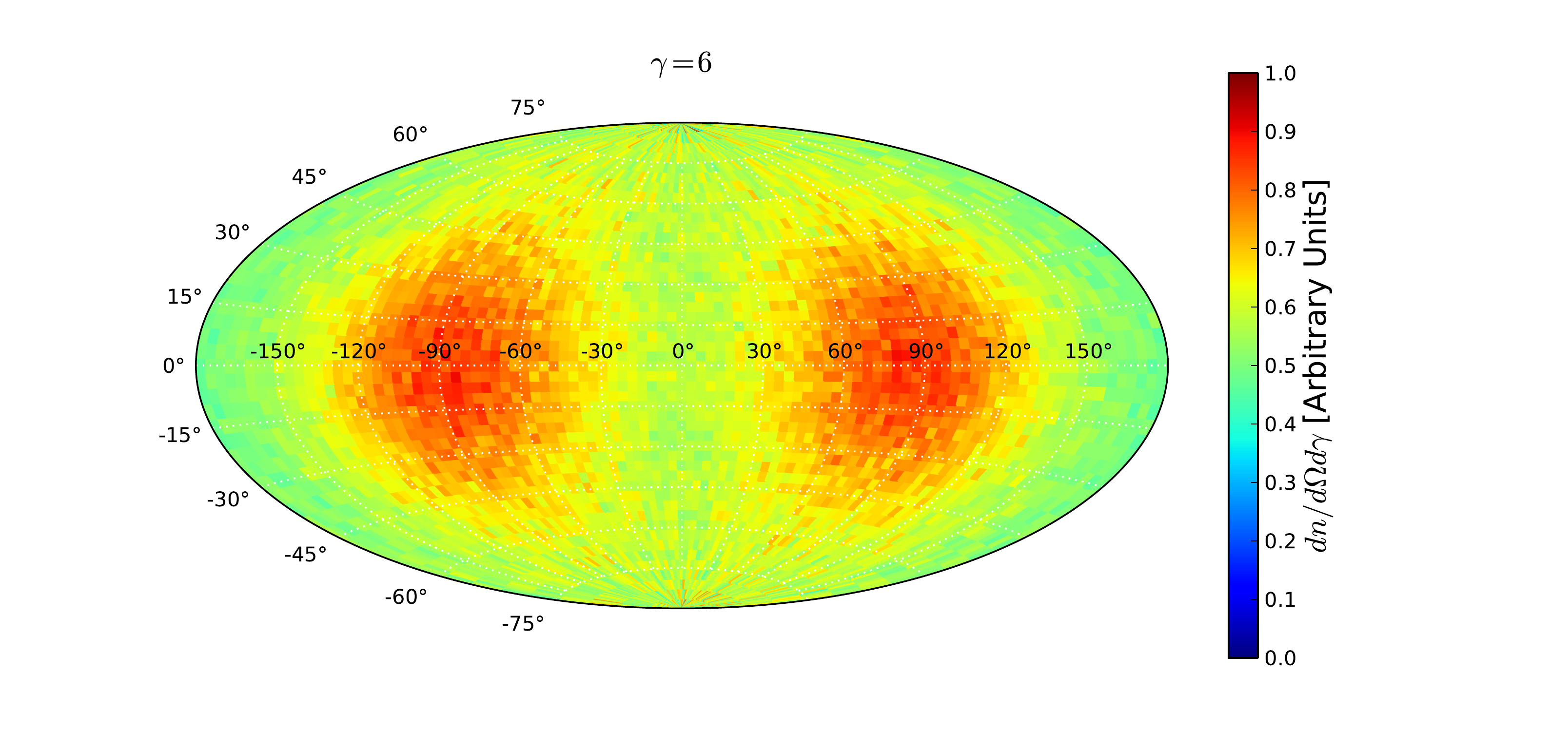}
\includegraphics[width=10cm]{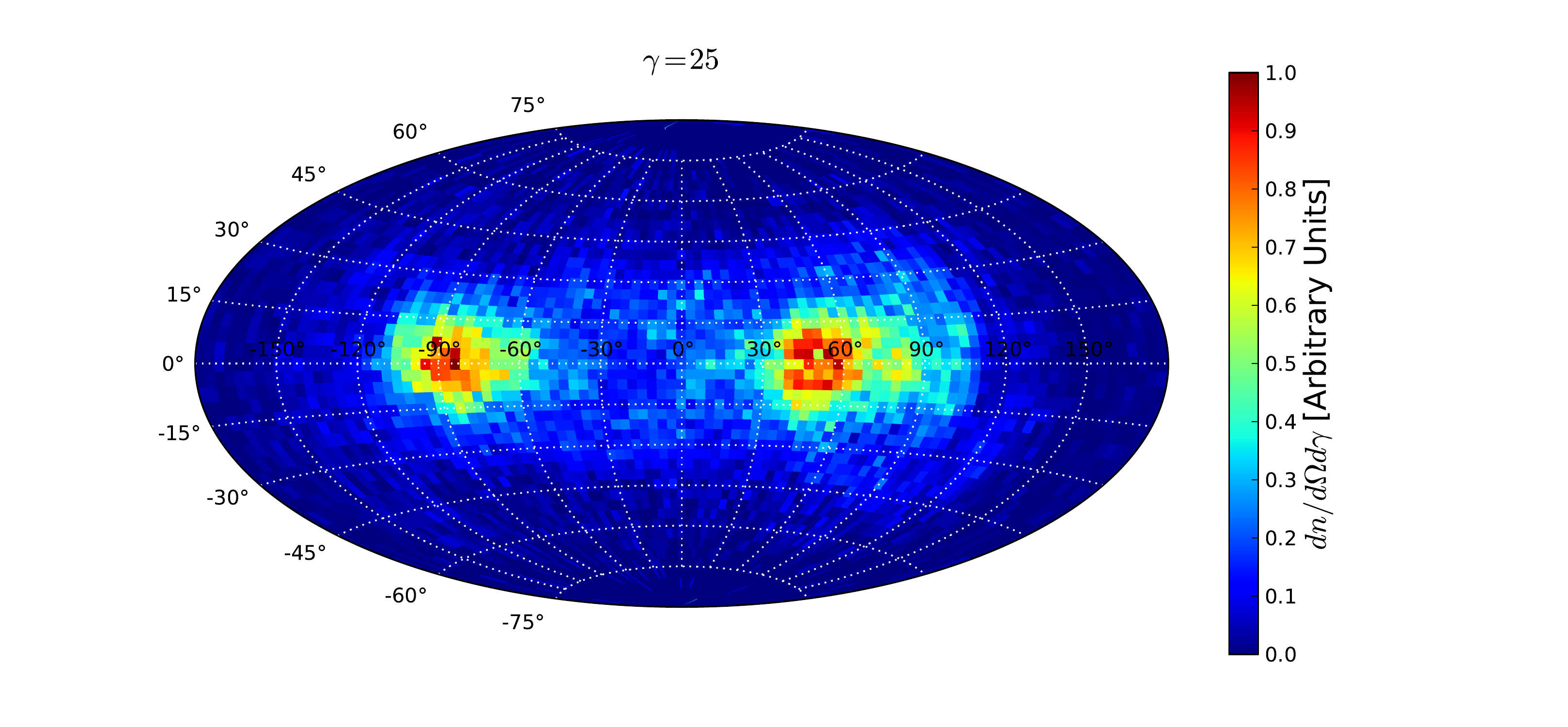}
\caption{Angular distribution of the particle 4-velocity vectors $\mathbf{u}$, $d{\rm n}/d\Omega d\gamma$ (contour plot), in three energy bins: $\mathbf{\gamma=1.5\pm 0.1}$ (top), $\mathbf{\gamma=6\pm 0.3}$ (middle), and $\mathbf{\gamma=25\pm 1.2}$ (bottom). In this projection (Aitoff), each direction is given by the latitude angle ($\sin\phi=u_{\rm y}/|u|$ with $-90^{\circ}<\phi<+90^{\circ}$, vertical axis) and the longitude angle ($\cos\lambda=u_{\rm z}/\sqrt{u_{\rm x}^2+u_{\rm z}^2}$ with $-180^{\circ}<\lambda<+180^{\circ}$, horizontal axis).  The precise geometry of the simulation is shown in Fig.~\ref{fig_bulk1}. These results were obtained from a 2D PIC simulation with $\sigma=10$ with no guide field (see also \citealt{2012ApJ...754L..33C, 2013ApJ...770..147C}, and \citealt{2014ApJ...782..104C} in 3D).}
\label{fig_anis}
\end{figure}

\begin{figure}[]
\centering
\includegraphics[width=13.5cm]{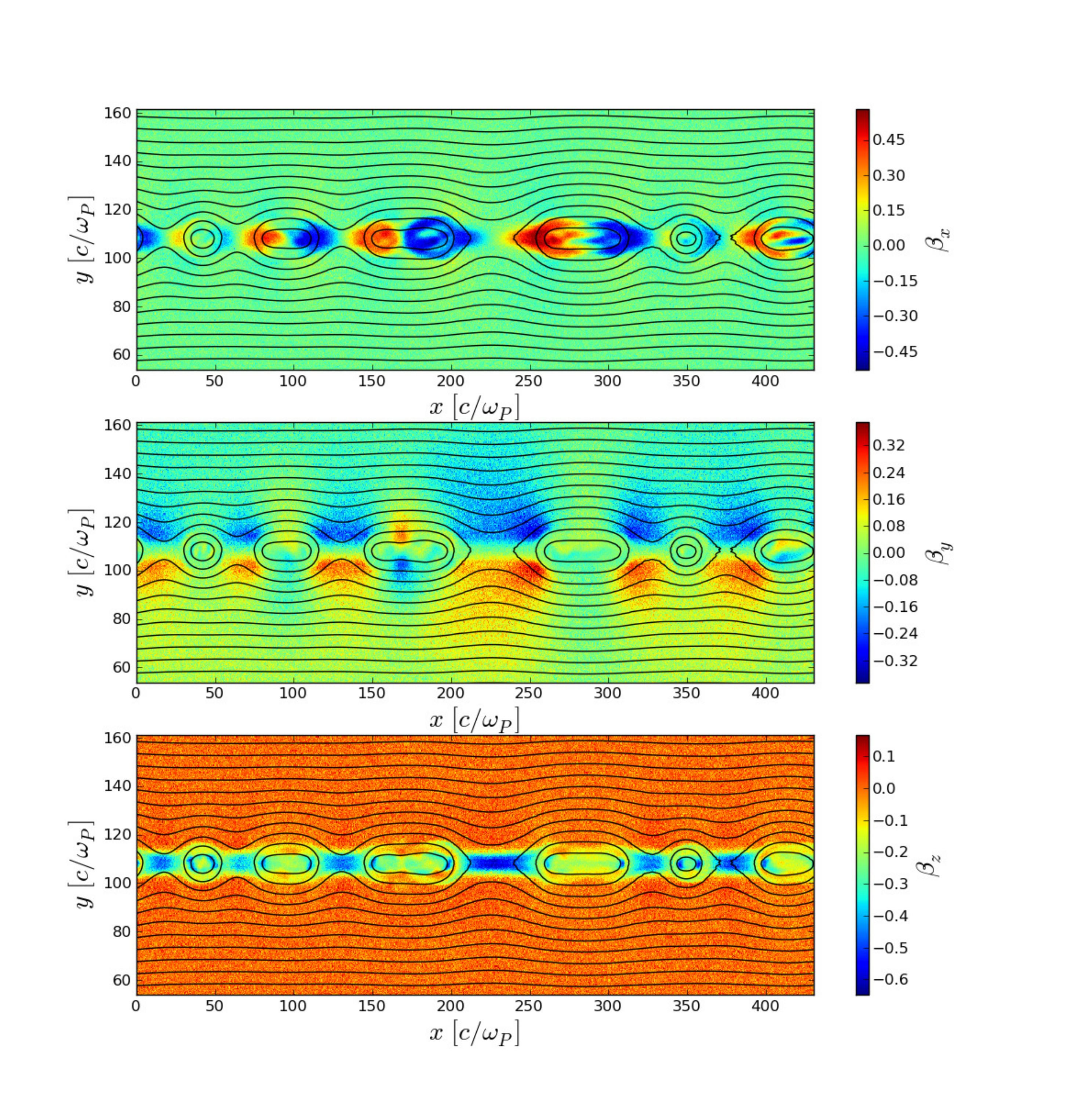}
\caption{Positron fluid velocity $\boldsymbol{\beta}=\mathbf{v}/c$ in the $x$- (top), $y$- (middle), and $z$-directions (bottom), for the same simulation as in Fig.~\ref{fig_anis} ($\sigma=10$, no guide field). The black solid lines show the magnetic field lines. The electron fluid velocity maps are identical, except that $\beta_{\rm z,electrons}=-\beta_{\rm z,positrons}$.}
\label{fig_bulk1}
\end{figure}

Fig.~\ref{fig_anis} presents the angular distribution of the particle 4-velocity vectors  as a function of the particle energy, from a 2D PIC simulation with $\sigma \approx10$ and with no guide field as first reported by \citet{2012ApJ...754L..33C}. The low-energy particles ($\gamma\sim 1$, top panel) present little anisotropy because these particles have not been accelerated at X-points. At higher energies ($\gamma\gtrsim\sigma$, middle and bottom panel), the particles exhibit clear sign of anisotropy with two beams pointing roughly towards the $\pm x$-directions, i.e., along the reconnection exhausts. Hence, the beams are not necessarily pointing along the reconnection electric field because the tension of the reconnected field lines pushes the particles away from the X-points in the form of a reconnection outflow towards the magnetic islands (see \fig{fluid2d}a, and top panel in Fig.~\ref{fig_bulk1}). Nonetheless, the direction of the beam of energetic particles is not static: it wiggles rapidly within the $(xz)$-plane (along the horizontal axis in Fig.~\ref{fig_anis}), which results in rapid flares of energetic radiation when the beam crosses the line of sight of a distant observer \citep{2012ApJ...754L..33C}. This result has interesting application to astrophysical flares, and in particular to the recently discovered $>100$~MeV gamma-ray flares discovered in the Crab Nebula \citep{2013ApJ...770..147C, 2014ApJ...782..104C} (see Sect.~\ref{PWN}). The Crab flare case is quite extreme in the sense that the particles emitting $>100$~MeV synchrotron radiation should be accelerated and radiating over a sub-Larmor timescale, so the highest energy radiation should keep the imprint of the particle anisotropy (regardless of the acceleration process), while the low-energy radiation should be more isotropic.

\begin{figure}[tbp]
\begin{center}
\includegraphics[width=0.8\textwidth]{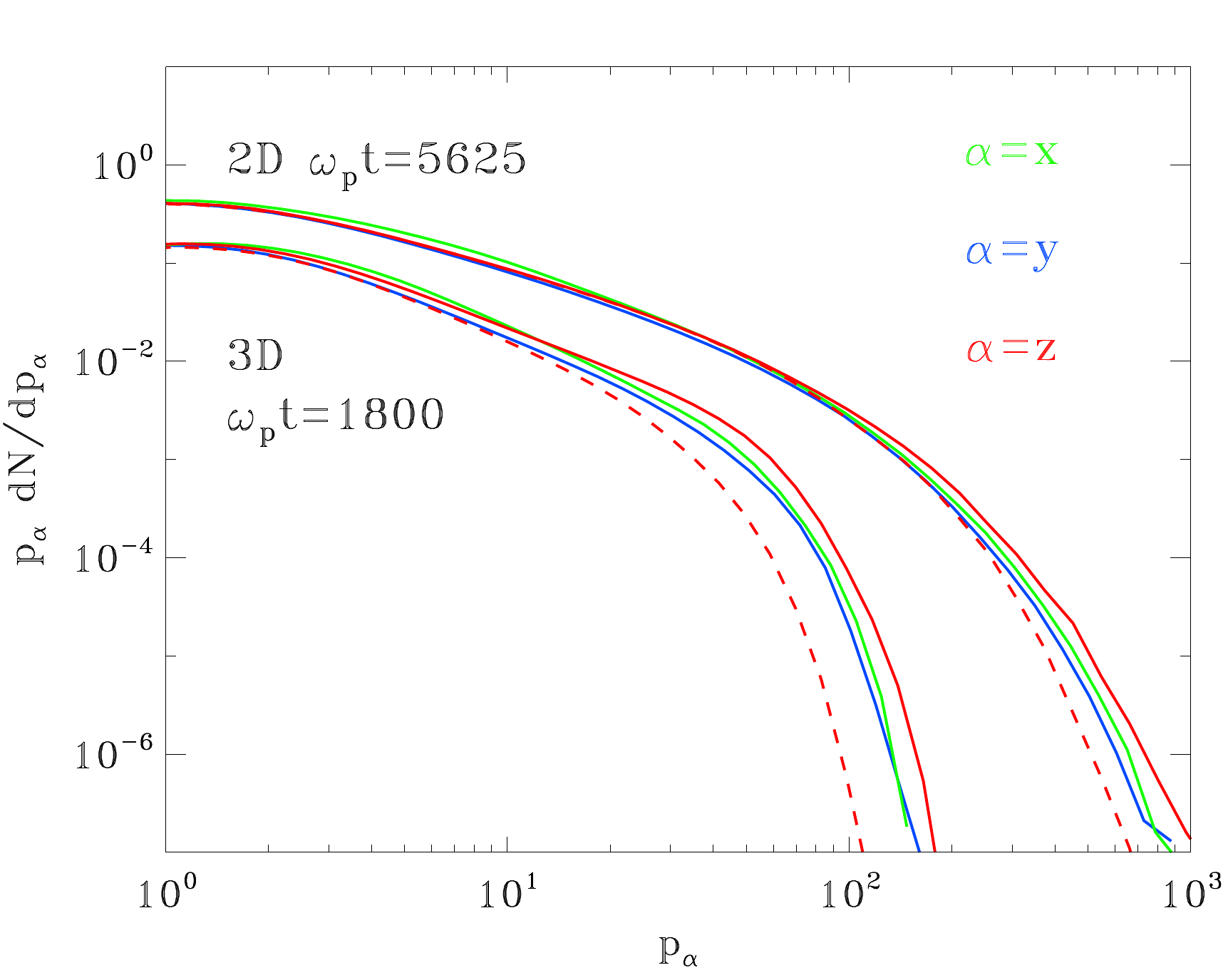}
\caption{Positron momentum spectrum along $x$ (green), $y$ (blue), $+z$ (red solid) and $-z$ (red dashed), for 2D and 3D, as indicated in the legend.}
\label{fig:spec3db}
\end{center}
\end{figure}

The pronounced anisotropy discussed above lasts for some limited amount of time. Indeed, when the high-energy particles reach the magnetic islands, they isotropize quickly in the strong fields shown in \fig{fluid2d}c and they do not contribute to the beamed emission. Since most of the particles at late times are contained in the major islands, it is not surprising that the long-term momentum spectra show little signs of anisotropy (see \fig{spec3db}). Even the residual difference between the momentum spectra along $+z$ and $-z$ (red solid and dashed lines, respectively) diminishes at later times (the 2D momentum spectra at $\ompt=1800$ were similar to the 3D results in \fig{spec3db}, showing that the anisotropy decays over time).

\begin{figure}[]
\centering
\includegraphics[width=13.5cm]{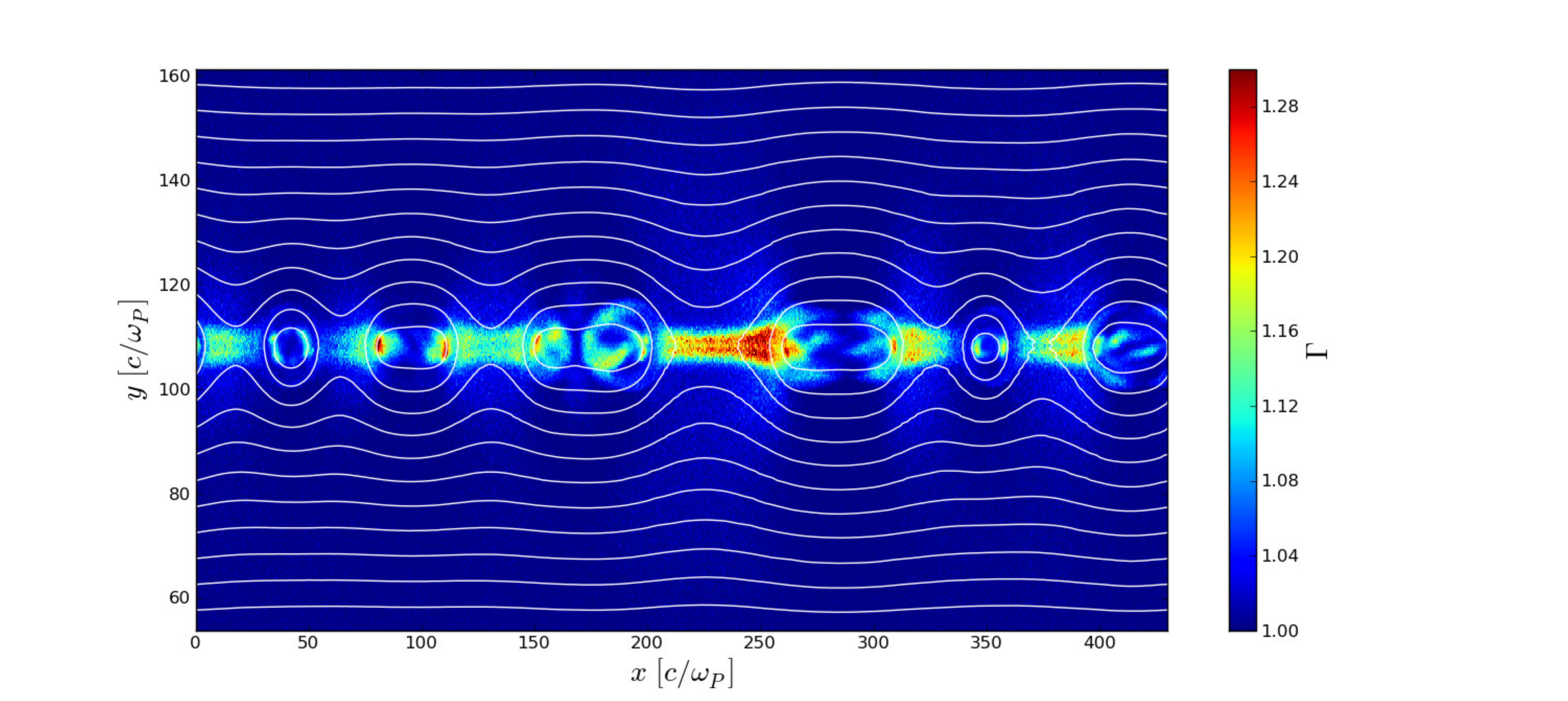}
\caption{Total Lorentz factor of the positron fluid, $\Gamma=1/\sqrt{1-\beta^2}$, computed from Fig.~\ref{fig_bulk1}. The white solid lines are magnetic field lines.}
\label{fig_bulk2}
\end{figure}

It is important to stress that this beaming mechanism is strongly energy-dependent. It should be distinguished from the Doppler boosting due to a relativistic bulk motion in the flow which beams all the particles and radiation by the same factor. In fact, relativistic reconnection produces also relativistic bulk flows as anticipated by \citet{2005MNRAS.358..113L}, and constitutes the cornerstone of the fast-variability models for blazar jets by \citet{2009MNRAS.395L..29G} (see Sect.~\ref{AGN}). Fig.~\ref{fig_bulk1} shows the three components of the fluid velocity vector normalized by the speed of light, $\boldsymbol{\beta}=\mathbf{v}/c$, for the same simulation as in Fig.~\ref{fig_anis} (where $\sigma=10$ and with no guide field) and at the same stage. The $x$-component presents the characteristic signature of a dipolar relativistic flow at every X-point where $\beta_{\rm x}\approx \pm 0.5$, which corresponds to the reconnection outflow accelerated by the tension of the newly reconnected field lines (i.e., $v_{\rm out}/c$ defined in Sect.~\ref{models}). The $y$-component shows the inflow of particles from the upstream towards the X-point that feeds the reconnection process with fresh plasma (i.e., $v_{\rm in}/c$ in Sect.~\ref{models}). This motion is due to the $\mathbf{E_{\rm z}}\times \mathbf{B_{\rm x}}$ drift velocity, and is about $\beta_{\rm y}\approx\pm 0.3$ in this particular simulation. The $z$-component is related to the electric current carried by counter-streaming electrons and positrons around the X-points. The corresponding fluid velocity is about $\beta_{\rm z}=\beta_+=-0.6$ for the positrons and $\beta_{\rm z}=\beta_-=+0.6$ for the electrons, but the net velocity is close to zero if both fluids are combined. Overall, the bulk Lorentz factor is only close to unity in this simulation (see Fig.~\ref{fig_bulk2}), which demonstrates that the anisotropic particle distributions is not related to the relativistic Doppler beaming. This being said, according to \citet{2005MNRAS.358..113L}, the bulk Lorentz factor of the outflow in relativistic Petscheck-like reconnection should scale as $\gamma_{\rm out}\sim\sqrt{\sigma}$. Indeed, it is hard to envision a scenario of fast reconnection (in the high $\sigma$ regime) where the outflowing material is not in relativistic bulk motion. PIC simulation runs that follow the evolution of the current sheet on a longer time scale typically find that the $\gamma_{\rm out}\sim\sqrt{\sigma}$ scaling works in the high-$\sigma$ regime (\citealt{2014ApJ...783L..21S}, K.~Nalewajko 2013, private communication).

\section{Astrophysical applications}\label{applications}

\subsection{Pulsars and pulsar wind nebulae}\label{PWN}

Pulsars are often regarded as one of the most suitable astrophysical environment for relativistic pair plasma reconnection. These objects are known to generate extremely magnetized plasma of pairs within their co-rotating magnetosphere. The plasma is released in the form of a relativistic magnetized wind beyond the light-cylinder surface, which is defined where the co-rotating velocity with the star equals the speed of light. In the wind region, the magnetic field lines open up and become mostly toroidal due to the fast rotation of the neutron star. This configuration naturally results in the formation of an equatorial current sheet (or ``striped wind'') that separates the two magnetic polarities.  This is the relativistic analog of the well-known ballerina's skirt shaped heliospheric current sheet.

Reconnection in the equatorial current sheet was first proposed by \citet{1990ApJ...349..538C} and \citet{1994ApJ...431..397M} as a remedy to the ``sigma-problem'', i.e., to explain the transition between a Poynting-flux dominated flow formed close to the neutron star ($\sigma\gg 1$) to the observed low-$\sigma$ pulsar wind nebulae. However, \citet{2001ApJ...547..437L} noticed that the dissipation of the current sheet would be followed by the acceleration of the wind. In the Crab pulsar, the wind would reach the termination shock before reconnection could proceed, unless the pulsar injects pairs at a higher rate than usually expected \citep{2003ApJ...591..366K}. As an alternative to the classical magnetospheric models (e.g., polar-cap, outer-gap, slot-gap), \citet{1996A&A...311..172L} suggested that reconnection in the striped wind could also explain the high-energy gamma-ray emission observed in pulsars \citep{2002A&A...388L..29K, 2012MNRAS.424.2023P, 2013A&A...550A.101A, 2014ApJ...780....3U}.

If, however, reconnection is inefficient in the wind zone, the striped wind is forced to dissipate at the termination shock \citep{2003MNRAS.345..153L}. Using particle-in-cell simulations, \citet{2007A&A...473..683P} in 1D and \citet{2011ApJ...741...39S} in 2D and 3D showed that shock-driven reconnection is able to annihilate the magnetic structure and efficiently accelerates particles regardless of the wind properties for large magnetizations. Whether the dissipation occurs in the wind or at the termination shock, it solves only partially the sigma-problem because the striped wind covers only a fraction of the solid angle set by the inclination angle between the rotation axis and the magnetic axis. Hence, the wind and the nebula should remain magnetically dominated at high latitudes (except for an orthogonal rotator). But, as we know from observations, pulsar wind nebulae are particle kinetic energy dominated flows, so there must be an extra mechanism to dissipate the remaining Poynting flux. \citet{1992SvAL...18..356L} and \citet{1998ApJ...493..291B} argued that pulsar wind nebulae should be subject to non-axisymmetric kink-like instabilities. Their hypothesis was recently corroborated by 3D relativistic MHD simulations by \citet{2011ApJ...728...90M} and \citet{2013MNRAS.431L..48P, 2014MNRAS.438..278P}. The dissipation of the magnetic energy could be done during the non-linear development of these instabilities via non-ideal MHD effects such as magnetic reconnection.

\begin{figure}[]
\centering
\includegraphics[width=12cm]{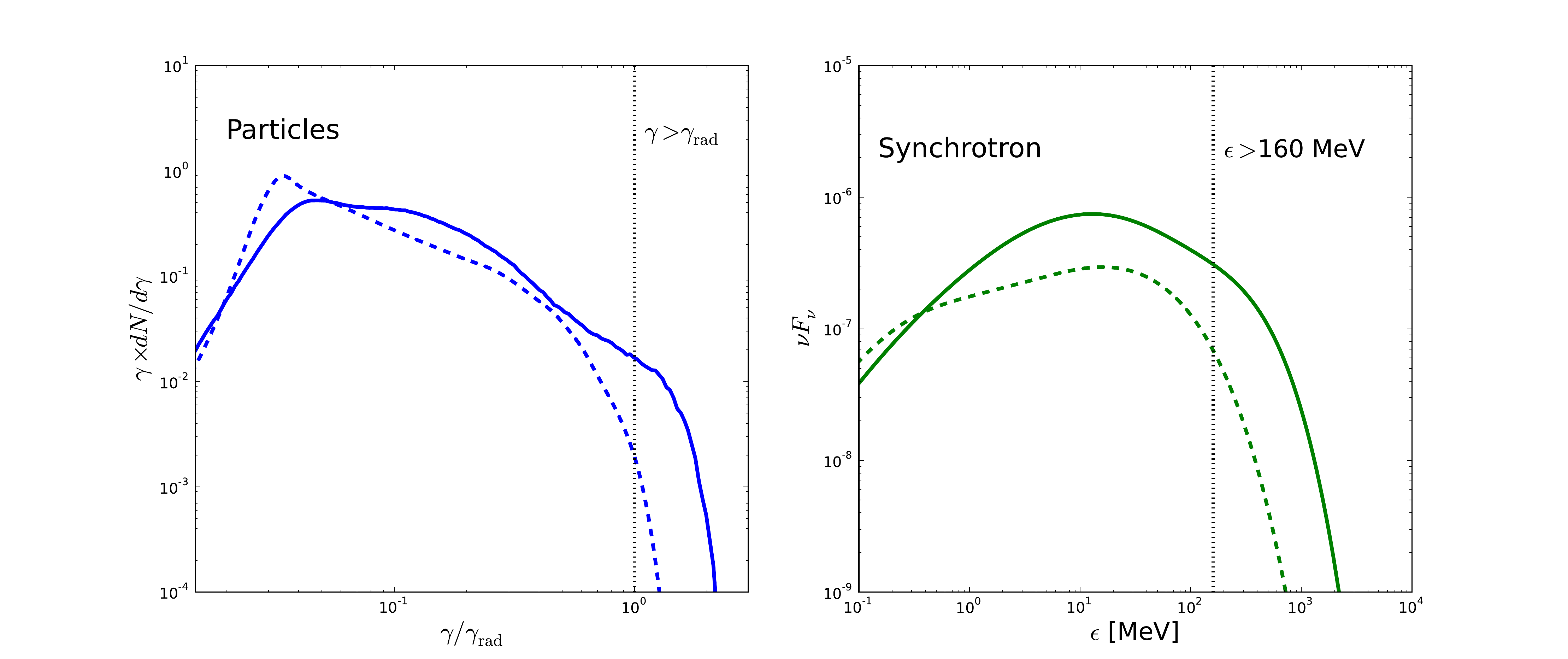}
\caption{Isotropically-averaged particle spectrum ($\gamma d{\rm N}/d\gamma$, left panel) and synchrotron radiation energy distribution ($\nu F_{\nu}$, right panel) in a 2D (solid line) and 3D (dashed line) PIC simulations of relativistic reconnection, including the effect of the radiation reaction force on the particles. The vertical dotted lines show the radiation-reaction limited energy of a particle if $E=B_0$ ($\gamma=\gamma_{\rm rad}$, left), and the corresponding maximum synchrotron photon energy ($\epsilon=160~$MeV independent of $E$ and $B_0$, right). Figure adapted from \citet{2014PhPl...21e6501C}.}
\label{fig_burnoff}
\end{figure}

The surprising discovery of short-lived, bright gamma-ray flares from the Crab Nebula \citep{2011Sci...331..736T, 2011Sci...331..739A} could be the direct evidence of magnetic reconnection in the Nebula \citep{2011ApJ...737L..40U, 2012MNRAS.426.1374C, 2014PhPl...21e6501C}. Using 2D and 3D PIC simulations, \citet{2013ApJ...770..147C, 2014ApJ...782..104C} showed that most of the features of the flares can be explained with relativistic reconnection (timescale, energetics, particle and photon spectra). In particular, these studies demonstrated that reconnection can accelerate particles above the synchrotron radiation burn-off limit \citep{1983MNRAS.205..593G, 1996ApJ...457..253D} deep inside the reconnection layer where the electric field overcome the magnetic field (see Fig.~\ref{fig_burnoff}), as anticipated by \citet{2004PhRvL..92r1101K} and \citet{2007A&A...472..219C} (Sect.~\ref{intro_rad}). This result is crucial because it can explain the emission of $>100~$MeV synchrotron radiation emitted during every Crab flare, which would be impossible to achieve in ideal MHD. The reconnection scenario would work best in the most magnetized regions of the nebula, i.e., near the poles and possibly in the jets \citep{2012ApJ...746..148C, 2012MNRAS.427.1497L, 2013MNRAS.428.2459K, 2013MNRAS.436.1102M}. Unfortunately, the current gamma-ray telescopes do not have the angular resolution to pin down the precise location of the flares within the Nebula.

\subsection{Jets from Active Galactic Nucleii}\label{AGN}

Jets from active-galactic nuclei have been monitored for decades at practically all accessible electromagnetic wavelengths resulting in a very rich phenomenology \citep{1995PASP..107..803U}. When the jet is pointing close to our line of sight, it is referred to as a ``blazar''. Recent observational progress in the blazar field has been immense. In particular, Cherenkov telescopes can now detect minute timescale variability in an increasing number of blazars \citep{2007ApJ...664L..71A}. These novel results strongly constrain the hydrodynamical models for the jet emission.

A broader consensus has emerged regarding the qualitative nature of the ``central engine''. The energy source in this view is a spinning black hole or the inner accretion disk threaded by a strong magnetic field (see, e.g., \citealt{1982MNRAS.199..883B}). This field transfers rotation energy outward as a Poynting flux. While part of the magnetic energy is used for the bulk acceleration of the jet, much of the energy remains in the magnetic field \citep{2010MNRAS.402..353L} and is available to power the jet emission through dissipation by instabilities and magnetic reconnection \citep{2009MNRAS.395L..29G}. In this picture, the jet is expected to be magnetically dominated in the emitting region, i.e., one deals with relativistic reconnection. In the following we show that applying our current understanding of relativistic reconnection to the physical conditions expected is blazar jets, reconnection can account for the extreme energetics and timescales inferred by blazar observations (for a similar approach to the modeling of the emission from gamma-ray bursts see \citealt{2005A&A...430....1G, 2006NJPh....8..119L, 2011ApJ...726...90Z}). The possibility that ultra-high-energy cosmic ray acceleration takes place at the current sheets
of the reconnection regions of powerful jets is investigated in \citet{2010MNRAS.408L..46G}.

\paragraph{The magnetic reconnection model for blazar emission:} Blazar emission varies on time\-scales typically ranging from hours to years and is thought to reflect, in part, variations of the gas properties in the black-hole vicinity\footnote{Several hours is the event-horizon light-crossing time of a billion solar-mass black hole-- mass typically inferred for the central engine in blazars: $t_{\rm cross}=2GM_{\rm BH}/c^3\simeq 10^4 M_9$ s.}. The recently discovered ultra-fast TeV flares from several blazars\footnote{Including Mrk 421, Mrk 501, PKS 2155-304, PKS 1222-216, and BL Lac.} (see, e.g., \citealt{2007ApJ...664L..71A, 2007ApJ...669..862A}) are strongly challenging the models for the blazar emission (\citealt{2008MNRAS.386L..28G,2009MNRAS.395L..29G}). This rare but generic blazar activity has several very revealing properties. (i) Fast flares have $\sim 10$ minute variability timescale, i.e, a factor $\sim 100$ shorter than the light-crossing time of the size of the black hole, pointing to extremely compact emitting regions. (ii) The emitting material must move with $\Gamma_{\rm em}\gtrsim 50-100$ for the TeV radiation to avoid anihhilation by soft radiation fields at the source \citep{2008MNRAS.384L..19B, 2008ApJ...686..181F}; these values of
$\Gamma_{\rm em}$ are much larger than the bulk jet motion $\Gamma_j\sim 10$ typically inferred in blazars from radio observations (see \citealt{2009AJ....138.1874L}). (iii) For $\gtrsim 100$ GeV photons to escape the {\it observed} broad line region of the blazar PKS 1222-216, the emitting region must be located at scales $\gtrsim 0.5$ pc \citep{2011A&A...534A..86T}. (iv) Simultaneous TeV and GeV ({\it Fermi-LAT}) observations indicate that the TeV flaring takes place on top of longer day-long blazar activity (e.g. \citealt{2011ApJ...733...19T}). (v) Fast flares may come in a repetitive fashion of similar events as observed in PKS 2155-304 \citep{2007ApJ...664L..71A}. Taken together, these inferences are extremely constraining for the models for the blazar emission.

\citet{2009MNRAS.395L..29G} argued that the ultra-fast variability must be generated internally in the jet by MHD instabilities. In strongly magnetized jets, the reconnection process injects energetic particles in compact, fast moving regions. These regions are natural emitters of powerful flares. Furthermore, the emitting material is expected to be faster than the jet on average allowing for TeVs to escape the source. 
For a jet moving with bulk $\Gamma_{\rm j}\sim 10-20$ and a plasmoid being ejected with  bulk 
$\gamma_{\rm out}\simeq \sqrt{\sigma}$ (as measured in the  rest frame of the jet), the
emitting region moves with $\Gamma_{\rm em}\simeq 2 \Gamma_{\rm
  j} \gamma_{\rm out}$ (in the frame of the host galaxy). For $\sigma\sim$ several, one can easily account for 
the required $\Gamma_{\rm em}\gtrsim 50$.  Applications of the model to fit spectra of specific sources are reported in \citet{2010MNRAS.402.1649G,2011MNRAS.413..333N}.

\begin{figure}[t]
\includegraphics[width=6cm]{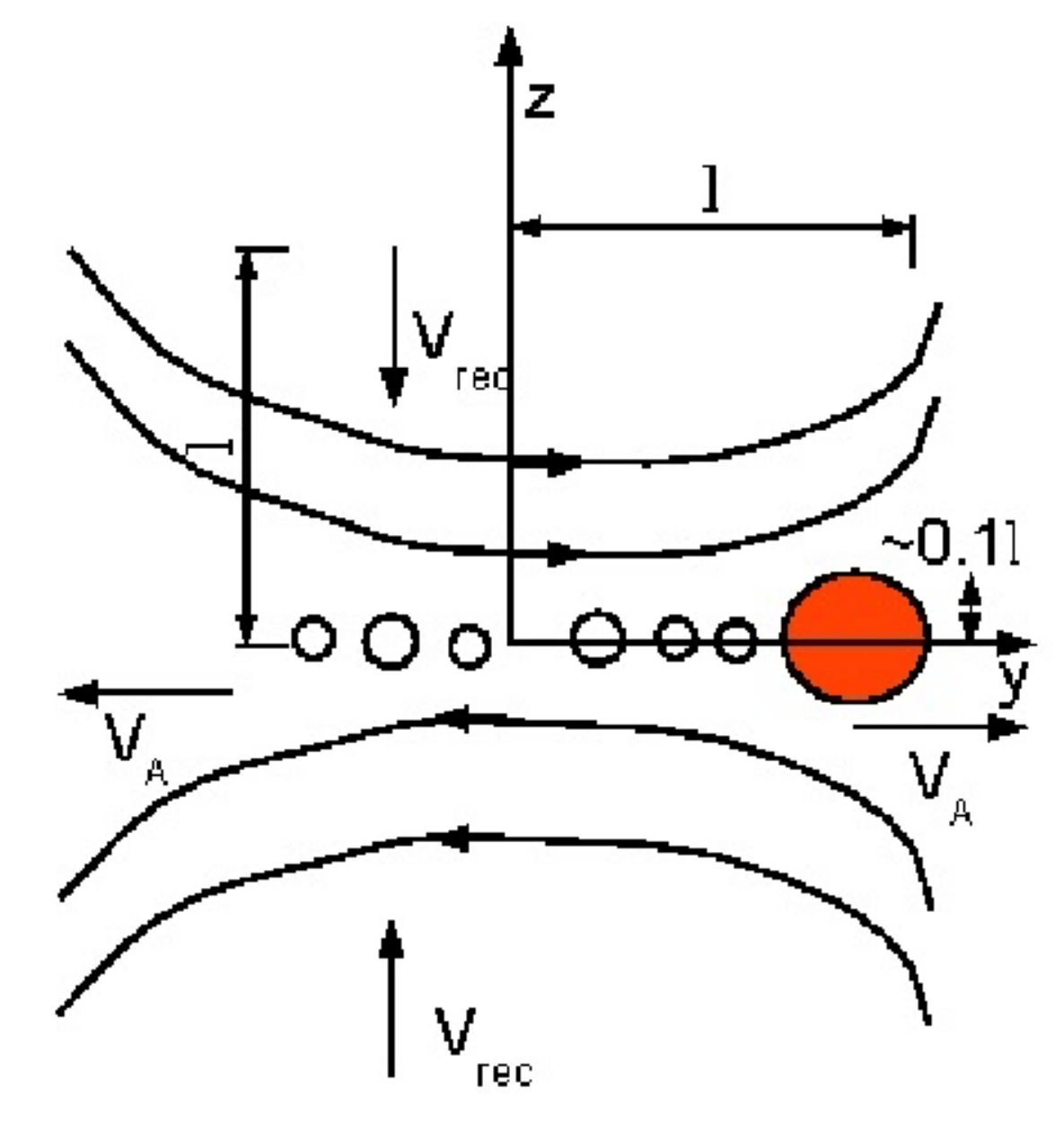}
\includegraphics[width=6cm]{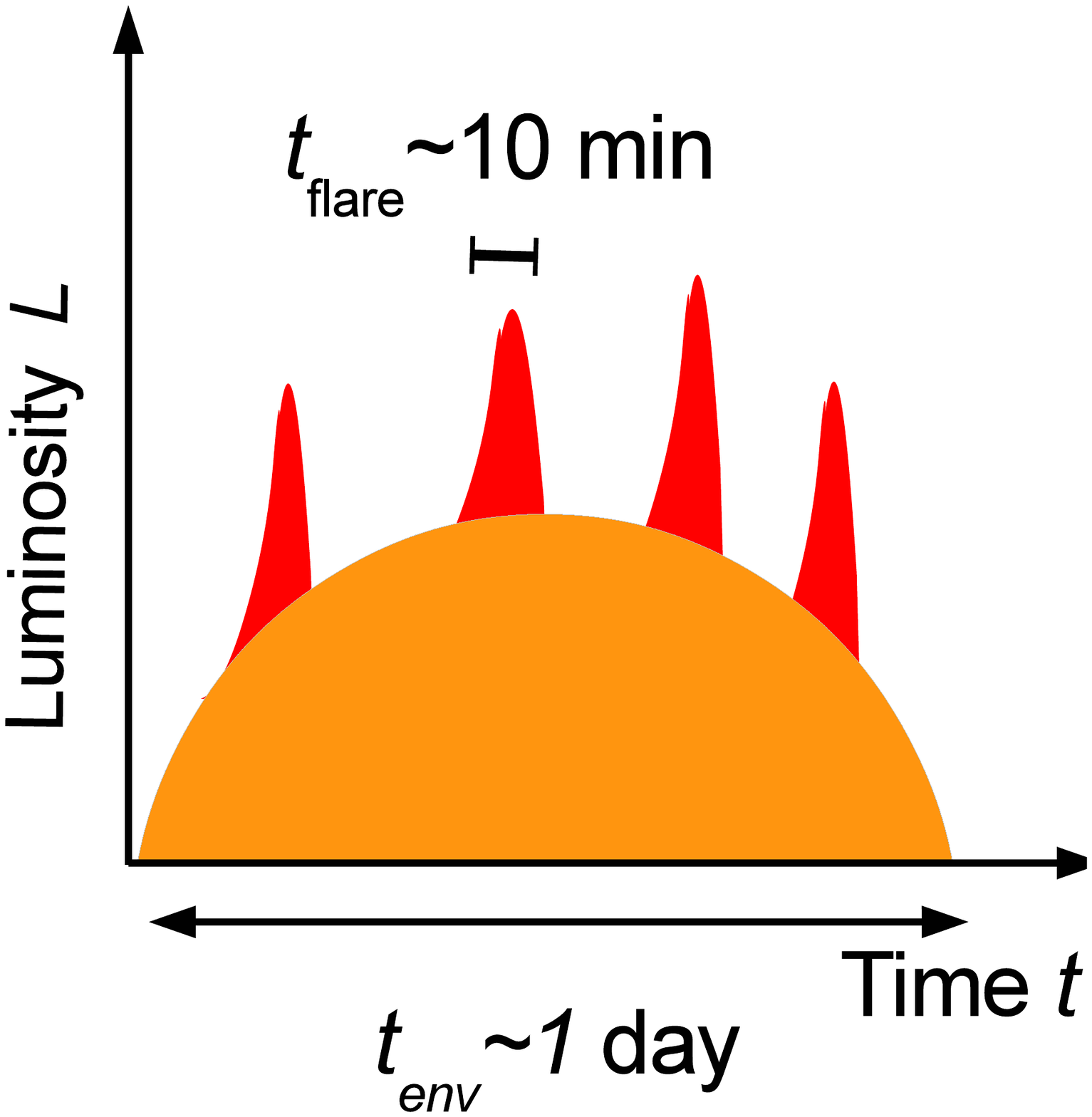}
\caption{{\it  Left Panel}: Schematic representation of the geometry of reconnection process shown in a frame comoving with the jet. Magnetic field lines of opposite polarity annihilate at the $x-y$ plane with speed $v_{\rm rec}=\beta_{\rm in} c$. The reconnection layer fragments to a large number of plasmoids. Regularily, plasmoids undergo multiple mergers resulting in a ``monster'' plasmoid (shaded blob). {\it Right Panel}: Sketch of the emission from plasmoid-dominated reconnection. The reconnection proceeds on a global timescale $t_{\rm rec}=l/\beta_{\rm in} c$, powering $\sim 1$day long flares (or envelope emission). Regularily, plasmoids grow to become ``monster'' plasmoids (shaded blob) giving rise to powerful, fast-evolving flares of duration $t_{\rm flare}\sim 10$ minutes. Several fast flares are expected from a single reconnection event.}
\label{fig:dimitrios}
\end{figure}

The \citet{2009MNRAS.395L..29G} model is based on a simplified picture for the reconnection geometry adopting a steady state reconnection model. As pointed out by \citet{2012MNRAS.420..604N} steady reconnection cannot account for the fastest evolving blazar flares because the variability timescale is limited by the reconnection speed $\beta_{\rm in}<1$.  
However, assuming steady reconnection is over-simplistic. Solar and Earth magnetosphere observations and recent advances in theory and numerical simulations (see previous Sections) have revealed that reconnection is an inherently time-dependent,
highly dynamic, process (see, e.g., \citealt{2005ApJ...622.1251L, 2006PhRvL..96s5003P, 2010SoPh..266...71K}). These time-dependent aspects of reconnection are crucial in understanding the fastest timescales involved in blazar flaring. For the physical conditions prevailing in jets, the reconnection current sheets are expected to suffer from tearing instabilities that lead to their fragmentation to a large number of plasmoids \citep{2007PhPl...14j0703L, 2009PhPl...16k2102B}. The plasmoids grow rapidly through mergers before leaving the reconnection region. Occasionally plasmoids  undergo significant growth to a scale of order of that of the reconnection region, forming ``monster'' plasmoids (\citealt{2010PhRvL.105w5002U}; see Fig.~\ref{fig:dimitrios}; left panel). The relativistic motion of the plasmoids in the rest frame of the jet results in additional beaming of their emission (i.e., beyond that induced by the jet motion). When the layer's orientation is such that plasmoids beam their emission towards the observer, powerful and fast evolving flares emerge. {\it Here we focus on the characteristic observed timescales and luminosities resulting from plasmoids that form in the reconnection region.} For simplicity, we assume that the dissipated energy is efficiently converted into radiation.\footnote{In practice the blazar emission is likely to result of ultrarelativistic electrons cooling via synchrotron radiation and Compton scattering. As discussed in previous sections, relativistic reconnection is an effective means of accelerating particles to such extreme energies.}

\citet{2013MNRAS.431..355G} demonstrated that a broad range of blazar phenomenology can be qualitatively understood in the context of plasmoid-dominated reconnection. The virtue of the model is that it can be applied to all blazar sources with observed fast flaring for similar adopted parameters. The model favors pc-scale dissipation for the origin of the fast flaring and provides theoretical motivation for such dissipation distance. Another interesting aspect of the model is that a sequence of fast flares is expected to have similar timescale set by the size of the reconnection layer as observed in PKS 2155. This work has demonstrated that the tight energetic, emitter Lorentz factor, and timescale constraints (i)-(v) are satisfied in the reconnection model. {\it More importantly, the basic assumptions of the
Giannios 2013 analysis on the properties of the reconnection layer have been  fully
verified by PIC simulations since then (see previous Sections).}

{In the following of this Section we make a plausibility argument for the model: 
we estimate the characteristic observed timescales and luminosities
resulting from plasmoids that form in the reconnection region} (for
full derivations see Giannios 2013). To
this end, we consider a blob (or plasmoid) emerging from the reconnection layer 
moving with the Alfv\'en speed of the reconnection upstream, i.e, with a corresponding 
bulk Lorentz factor $\gamma_{\rm out}\simeq \sqrt{\sigma}$
(measured in the jet rest frame) and of size $R_{\rm p}''=fl'$, where $l'$ is the characteristic scale 
of the reconnection region and $f$ is a dimensionless parameter of
the order of 0.1, as expected for the largest, ``monster''  plasmoids (Uzdensky et
al. 2010); hereafter, primed
(double primed) quantities are measured in the rest frame of the jet
(emitting blob).\footnote{We assume that the plasmoid instability operates across the whole length of the
current sheet, as opposed to a situation where central, very compact, dissipation region
forms and is surrounded by extended magnetic separatrices (the slow shocks in Petscheck model)
across which most of the plasma flows. In the latter case, the monster plasmoids may be smaller.} The observed characteristic variability time for the plasmoid 
emission is $t_{\rm v}\simeq R''_{\rm p}/\delta_p c$, where $\delta_{\rm p}$ is the Doppler 
boost of the plasmoid radiation towards the observer. For a central engine in which the 
magnetic field varies on a dynamical time $\sim R_{\rm Sch}/c$, the characteristic scale of 
the reconnection region  can be estimated to be $l'\simeq\Gamma_{\rm j} R_{\rm Sch}$ resulting in
\begin{equation}
t_{\rm v}=\frac{f\Gamma_jR_{\rm Sch}}{\delta_{\rm p}c}=400
f_{-1}\Gamma_{\rm j, 20}M_9\delta_{p, 50}^{-1}\rm \quad s,
\label{eq:1}
\end{equation}
where $\delta_{\rm p}=50\delta_{\rm p,50}$, $f=0.1f_{-1}$, $\Gamma_{\rm j}=20\Gamma_{\rm j,20}$. $f\sim 0.1$ describes the largest plasmoids expected in the layer \citep{2010PhRvL.105w5002U}. Flaring on several minute timescale is therefore expected in this picture.

Consider a jet emerging from a supermassive black hole with (isotropic equivalent) power $L_{\rm iso}$, opening angle $\theta_j$ and Lorentz factor $\Gamma_j$. We also assume that $\theta_j \Gamma_j=0.2$ as indicated by observations \citep{2009A&A...507L..33P}. The typical bulk Lorentz factor of gamma-ray active blazars is $\Gamma_j\sim 10-20$ \citep{2010A&A...512A..24S, 2012ApJ...758...84P}. The energy density at the dissipation, or ``blazar'', zone is
\begin{equation}U'_{\rm j}=\frac{L_{\rm iso}}{4\pi (\theta_{\rm j} R_{\rm
    diss})^2\delta_{\rm j}^4c}.\end{equation} The dissipation distance $R_{\rm diss}$ is estimated
requiring that the reconnection proceeds within the expansion time of the jet ($R_{\rm diss}/\Gamma_jc\sim l'/\epsilon c$).

Pressure balance across the reconnection layer requires the energy density of the plasmoid to be similar to that of the jet $U''_p\sim U'_j$. Assuming efficient conversion of dissipated energy into radiation, the rest-frame luminosity of the plasmoid is thus $L_{\rm p,obs}=\delta_p^4L'' =\delta_p^4U_{\rm p}''4\pi R_{\rm p}''^2c$. Putting everything together, the observed luminosity of the plasmoid is \citep{2013MNRAS.431..355G}
\begin{equation}
L_{\rm p,obs}=10^{47}\frac{\beta_{\rm in, -1}^2f_{-1}^2\delta_{p,50}^4L_{\rm iso,48}}{\delta_{j,20}^4}\quad \rm erg/s.
\label{eq:3}
\end{equation}

The Doppler factor of the plasmoid $\delta_{\rm p}$ depends on several parameters. It is related to $\Gamma_j$, $\gamma_{\rm out}$, the angle of the plasmoid with respect to the jet motion and the observer's angle of sight. For typical situations where the reconnection layer is at a large $\theta\sim \pi/2$ angle with respect to the jet propagation (as seen in the jet rest) and fairly aligned with the observer (giving powerful flares) $\delta_{\rm p} \sim \Gamma_{\rm j}\gamma_{\rm out}$. {One can see (see Eq.~\ref{eq:3}) that powerful flares on a timescale of $\sim$10 min is possible even with very modest relativistic motions within the jet $\gamma_{\rm out}\sim 2$.

\paragraph{Ejection of multiple monster plasmoids:} During a reconnection event multiple monster plasmoids are expected to form. 2D simulations \citep{2012PhPl...19d2303L} indicate that monster plasmoids form every few Alfv\'en times $t_A$ or at a rate of $\sim 0.3t_A^{-1}$. It appears likely that 2D simulations underestimate the rate of formation of monster plasmoids. The actual rate may be higher when the 3D structure of the layer is considered \citep{2014ApJ...783L..21S}. If monster plasmoids emerge at a rate $\sim (0.3-3)t_A^{-1}$, some $(3-30)/\beta_{\rm in, -1}$ plasmoids are expected from a single reconnection layer powering multiple flares. A sketch of such pattern is shown in Fig.~\ref{fig:dimitrios}.

\paragraph{The ``envelope emission'' from the reconnection region:} The bulk motion of a monster plasmoid is expected to be similar to the speed of other structures (e.g. smaller plasmoids) leaving the reconnection region. When the plasmoid emission is beamed towards the observer (powering a fast flare), the overall emission from the current layer is also beamed by a similar factor. The emission from the layer forms a slower-evolving ``envelope''. In the following we estimate the timescale and luminosity of the emission from the reconnection layer.

At the dissipation distance $R_{\rm diss}$, the reconnection proceeds within the expansion time of the jet ($R_{\rm diss}/\Gamma_jc\sim l'/\beta_{\rm in} c$) which is observed to last for $t_{\rm exp,obs}\simeq R_{\rm diss}/\Gamma_j^2c$. Therefore, $t_{\rm exp,obs}$ corresponds to the observed duration of the envelope emission which is simply (using
also Eq.~(\ref{eq:1})):
\begin{equation}
t_{\rm env}=\frac{R_{\rm diss}}{\Gamma_j^2c}=10^5\frac{M_9}{\beta_{\rm in, -1}} \quad \rm s. 
\end{equation}
The duration of the envelope emission is $\sim$days. Such timescale is characteristic of blazar flares.

The (lab frame) energy available to power the envelope emission is $E_{\rm env}=U_{\rm j}2l'^3/\Gamma_{\rm j}$, where $U_j=\Gamma_j^2U'_j$ is the energy density of the jet and $2l'^3/\Gamma_{\rm j}$ accounts for (lab frame) volume of the reconnection region that powers each minijet (see Fig.~\ref{fig:dimitrios}). The emitted luminosity of the reconnection region is $E_{\rm env}/t_{\rm env}$. It can be converted into {\it observed} luminosity by accounting for beaming factor of the emission $\sim \delta_p^2$:
\begin{equation}
L_{\rm env,obs}\simeq 2\Gamma_{\rm j}^2\delta_{\rm p}^2l'^2U_{\rm
  j}'\beta_{\rm in} c=3\times 10^{46}\frac{\Gamma_{\rm j,20}^2\delta_{\rm p,50}^2
\beta^3_{\rm in, -1}L_{\rm iso, 48}}{\delta_{j,20}^4} \quad \rm erg/s.
\label{eq:4}
\end{equation}

The envelope emission is quite bright. Dividing Eqs.~(\ref{eq:3}) and (\ref{eq:4}), one arrives to a fairly simple expression for the ratio of the plasmoid to envelope luminosities $L_p/L_{\rm env}\sim 3f_{-1}^2\delta_{\rm p,50}^2/(\Gamma_{\rm j,20}^2\beta_{\rm in, -1})$. The luminosity contrast depends only on the Lorentz factor of the minijet in the rest frame of the jet  $\gamma_{\rm p}\simeq \delta_{\rm p}/\Gamma_{\rm j}$, the size of the plasmoid parametrized by $f$, and the reconnection sped $\beta_{\rm in}$. The observed luminosity ratio is of order unity constraining $\delta_{\rm p,50}/\Gamma_{\rm j,20} \sim 1$ for $\beta_{\rm in}\sim f\sim 0.1$. The ratio $\delta_{\rm p,50}/\Gamma_{\rm j,20}$ is determined by the reconnection-induced bulk motions in the jet and points to $\gamma_{\rm out}\sim 2$ or, equivalently, moderately magnetized jet with $\sigma\sim $ several.

Most of the current numerical work on relativistic reconnection (and this review so far)
has focused on the case of electron-positron plasmas.
The composition of the jet flow is still an open question but an
electron-proton jet is a strong possibility. 
Electron-ion reconnection is more challenging, on a numerical level, 
than electron-positron reconnection, since the computation has to
resolve the small scales of electrons, yet the system evolves on the 
longer ion timescales.  However, the physics of relativistic electron-proton  reconnection, yet still at an early stage of investigation, shows remarkable similarities with electron-positron reconnection \citep[e.g.,][]{melzani14}.
A detailed investigation of relativistic reconnection in the case of
unequal mass charges is of paramount importance to obtaining  
predictions for the acceleration of electrons and cosmic rays in blazar jets. 

\section{Conclusion}\label{conclusions}

There has been significant progress in our understanding of relativistic reconnection in recent years, thanks to both analytical works and numerical simulations. One important outcome is that plasma instabilities in current sheets play a crucial role in the dynamics of reconnection. In particular, the tearing instability which fragments the current sheet, leads to fast reconnection and efficient non-thermal particle acceleration. Particle-in-cell simulations are now large enough to unambiguously identify broad, hard power laws in the particle energy distributions (in the high-magnetization limit). The power-law index is typically harder than the universal $\sim -2$ index expected in shock acceleration. These impressive developments were also motivated by puzzling observations of high-energy phenomena in the Universe, especially flaring gamma-ray sources. Ultra-rapid gamma-ray flares discovered in the Crab Nebula and in several AGN jets are too fast and too bright to be explained by conventional models. Particle beaming and relativistic bulk motions associated with relativistic reconnection can alleviate these difficulties. We expect fast new developments in this field, with more applications to astrophysical objects.

\begin{acknowledgements}
We thank the referees for useful comments that helped to improve the manuscript. LS is supported by NASA through Einstein
Postdoctoral Fellowship grant number PF1-120090 awarded by the Chandra
X-ray Center, which is operated by the Smithsonian Astrophysical
Observatory for NASA under contract NAS8-03060. BC acknowledges support from the Lyman Spitzer Jr. Fellowship awarded by the Department of Astrophysical Sciences at Princeton University, and the Max-Planck/Princeton Center for Plasma Physics. DG acknowledges support from the NASA grant NNX13AP13G. 
\end{acknowledgements}
\bibliographystyle{aps-nameyear}

\end{document}